\documentclass[sigplan,screen,pbalance]{acmart}

\AtBeginDocument{%
  }

\setcopyright{rightsretained}
\acmDOI{10.1145/3696443.3708948}
\acmYear{2025}
\copyrightyear{2025}
\acmISBN{979-8-4007-1275-3/25/03}
\acmConference[CGO '25]{Proceedings of the 23rd ACM/IEEE International Symposium on Code Generation and Optimization}{March 01--05, 2025}{Las Vegas, NV, USA}
\acmBooktitle{Proceedings of the 23rd ACM/IEEE International Symposium on Code Generation and Optimization (CGO '25), March 01--05, 2025, Las Vegas, NV, USA}
\received{2024-09-12}
\received[accepted]{2024-11-04}

\usepackage{listings}
\usepackage[frozencache,cachedir=.]{minted}
\usepackage{url}
\usepackage{comment}
\usepackage{hyperref}
\usepackage{multirow}
\usepackage{booktabs}
\usepackage{graphicx}
\usepackage{subfig}
\usepackage{float}
\usepackage{amsmath}
\usepackage{algorithm2e}
\RestyleAlgo{ruled}
\SetKwComment{Comment}{/* }{ */}
\definecolor{bg}{rgb}{0.95,0.95,0.95}
\newcommand{\omod}[1]{\ \mathrm{mod}\ #1}

\newcommand{\rewriteto}[0]{\quad\rightarrow\quad}
\usepackage{cuted}
\usepackage{scalerel}
\newcommand{\qeq}[0]{=_{\scaleto{?}{4pt}}\hspace{-2pt}}
\renewcommand{\paragraph}[1]{\textbf{#1}}
\usepackage{soul}

\begin{document}

\newcommand\naifeng[1]{\textcolor{cyan}{Naifeng: #1}}
\newcommand\franz[1]{\textcolor{red}{Franz: #1}}
\newcommand{\todo}{\textcolor{red}{[todo]}}

\newcommand{\DFT}{\operatorname{DFT}}
\newcommand{\NTT}{\operatorname{NTT}}
\newcommand{\stride}{\operatorname{L}}
\newcommand{\diag}{\operatorname{D}}
\newcommand{\tensor}{\otimes}
\newcommand{\dirsum}{\oplus}
\newcommand{\one}{\operatorname{I}}
\newcommand{\twiddle}{\operatorname{T}}
\newcommand{\rotation}{\operatorname{R}}

\title[Code Generation for Cryptographic Kernels on GPU]{Code Generation for Cryptographic Kernels using \\ Multi-word Modular Arithmetic on GPU}

\author{Naifeng Zhang}
\orcid{0009-0004-0190-4041}
\affiliation{%
  \institution{Carnegie Mellon University}
  \city{Pittsburgh}
  \country{USA}
}
\email{naifengz@cmu.edu}

\author{Franz Franchetti}
\orcid{0000-0002-3529-8973}
\affiliation{%
  \institution{Carnegie Mellon University}
  \city{Pittsburgh}
  \country{USA}
}
\email{franzf@andrew.cmu.edu}





\begin{abstract}

Fully homomorphic encryption (FHE) and zero-knowledge proofs (ZKPs) are emerging as solutions for data security in distributed environments. However, the widespread adoption of these encryption techniques is hindered by their significant computational overhead, primarily resulting from core cryptographic operations that involve large integer arithmetic. This paper presents a formalization of multi-word modular arithmetic (MoMA), which breaks down large bit-width integer arithmetic into operations on machine words. We further develop a rewrite system that implements MoMA through recursive rewriting of data types, designed for compatibility with compiler infrastructures and code generators.
We evaluate MoMA by generating cryptographic kernels, including basic linear algebra subprogram (BLAS) operations and the number theoretic transform (NTT), targeting various GPUs. Our MoMA-based BLAS operations outperform state-of-the-art multi-precision libraries by orders of magnitude, and MoMA-based NTTs achieve near-ASIC performance on commodity GPUs. 

\end{abstract}

\ccsdesc[500]{Theory of computation~Rewrite systems}
\ccsdesc[500]{Security and privacy~Cryptography}
\ccsdesc[500]{Computing methodologies~Parallel computing methodologies}


\keywords{Multi-word modular arithmetic, code generation, rewrite system, BLAS, number theoretic transform, cryptography}


\maketitle

\section{Introduction}

\begin{figure}[t]
\centering
    \vspace{1mm}
    \includegraphics[width=0.45\textwidth,trim={10mm 0mm 0mm 5mm},clip]{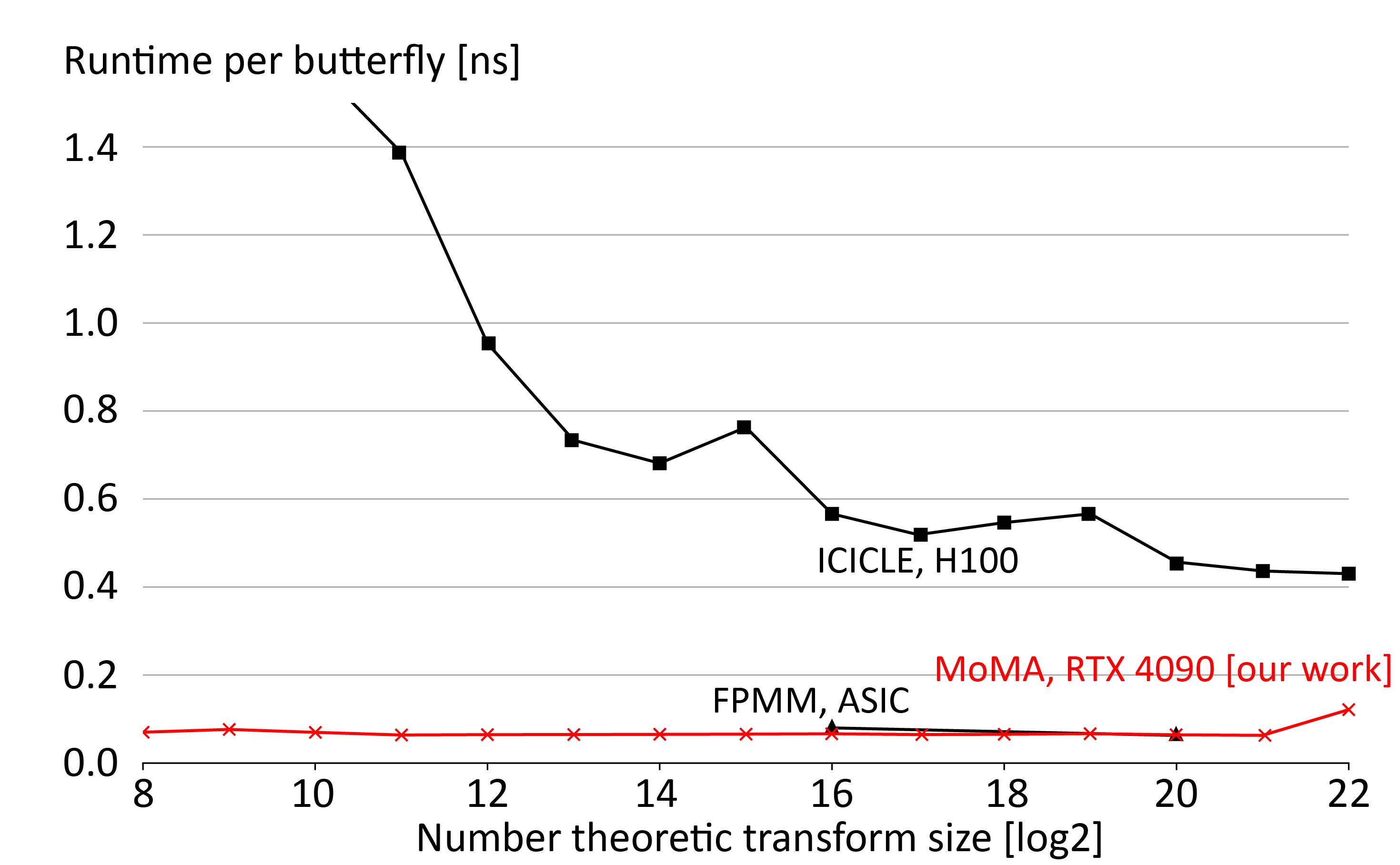}
    \caption{Performance of 256-bit NTT on GPUs and ASIC (lower is better). On NVIDIA GeForce RTX 4090, MoMA-based NTT outperforms state-of-the-art cryptographic acceleration library~\cite{inbasekar2024icicle} running on NVIDIA H100 by an average of 14 times and achieves near-ASIC~\cite{zhou2024fully} performance.}
    \label{fig:intro}
\end{figure}

As data security becomes increasingly critical in distributed and cloud computing environments, advanced encryption schemes such as fully homomorphic encryption (FHE) and zero-knowledge proofs (ZKPs) are emerging as promising solutions for privacy preservation. However, these encryption techniques remain impractical for widespread use due to their computational overhead, much of which originates from core cryptographic operations involving large integer arithmetic. These operations are typically performed on large finite fields, where integer bit-width ranges from 256 to 768 bits for ZKP applications and from hundreds to over 1,000 bits for FHE-based systems. 

To manage this computational cost, current ZKP implementations often rely on arbitrary-precision libraries such as the GNU Multiple Precision (GMP) library~\cite{granlund1996gnu} or programming languages such as Rust and Python, which support large integer arithmetic natively. However, while these external libraries offer fundamental support for large integer operations, they come with limitations. For instance, many ZKP libraries are written in languages like Python that are not performance-oriented, or are constrained to CPU execution due to factors such as GMP's lack of support for GPUs.
As integer bit-widths in popular FHE schemes extend to thousands of bits, the residue number system (RNS) is used to represent these large numbers via much smaller integers (residues) that fit within machine words, typically 32 or 64 bits. However, using RNS with small residues introduces additional computational overhead for modulus raising and reduction~\cite{cheon2019full} and requires frequent bootstrapping, which is a highly computationally intensive process in FHE~\cite{al2023demystifying}. 

To address these challenges, we formally define \emph{multi-word modular arithmetic} (MoMA), in which large integers are represented as multiple machine words to exploit the native performance of single-word arithmetic. While similar ideas have been explored in the past, such as Intel's work in the 1980s~\cite{guide2011intel}, to the best of our knowledge, our work provides a novel formalization of multi-word modular arithmetic that \emph{systematically} decomposes large integer arithmetic into machine word operations. This formalization directly enables many optimizations in symbolic space, including reducing redundant operations for inputs with non-power-of-two bit-widths as opposed to simply zero-padding the inputs.
Moreover, we introduce a program transformation pass that implements MoMA as a set of rewrite rules in a term rewriting system. This pass, integrable with compiler frameworks and code generators, operates on data types and recursively transforms computations involving large data types into equivalent sequences of operations on smaller data types, continuing until all data types used in the kernel are natively supported by the machine. 

In this paper, we focus on targeting GPUs due to their massive parallelism and high on-chip performance. We implement MoMA in SPIRAL~\cite{franchetti2018spiral}, a code generator that produces highly efficient implementations for various hardware architectures and provides strong support for implementing mathematically formal rule systems. 
To evaluate our approach, we first implement several basic linear algebra subprograms (BLAS) operations on finite fields, which correspond to polynomial arithmetic operations that are fundamental to many advanced cryptographic schemes~\cite{al2022openfhe}. Next, we implement a more complex kernel, the number theoretic transform (NTT), which is a critical component in FHE and ZKPs. NTTs account for over 90\% of the runtime in many FHE schemes and approximately 30\% in ZKP applications~\cite{fan2023tensorfhe,zhang2021pipezk}. Our MoMA-based implementation of BLAS operations outperforms state-of-the-art multi-precision libraries by orders of magnitude. Additionally, our MoMA-based NTT implementation achieves near-ASIC performance on a commodity GPU, as shown in Figure~\ref{fig:intro}, while maintaining the flexibility to support multiple input bit-widths across various NTT sizes.

By utilizing MoMA, we offer a more performant alternative to ZKP libraries that rely on GMP, Python or Rust for large integer arithmetic.
In the context of FHE, transitioning from 64-bit to 128-bit (and higher bit-width) residues in the RNS representation using MoMA creates opportunities to reduce the frequency of costly operations, such as bootstrapping. In some FHE schemes, MoMA could potentially eliminate the need for RNS entirely if the original bit-width is in the hundreds and the cost of using MoMA to decompose the integer is less than the cost of employing RNS. 
Historically, extending integer bit-width beyond the machine word width has been viewed as highly expensive in the field of cryptography. MoMA's ability to reduce the computational cost of large integer arithmetic could potentially enable innovative cryptographic algorithms that are not constrained by the limitations of 64-bit word width.

\paragraph{Contributions.} This paper makes the following contributions:
\begin{enumerate}
    \item A formal definition of \sloppy{multi-word modular arithmetic (MoMA)} that decomposes integer arithmetic on large bit-widths into native machine word operations.
    \item A rewrite system that implements MoMA, compatible with compiler frameworks and code generators, that recursively rewrites computations of large data types into an equivalent sequence of operations that use smaller data types. 
    \item A demonstration of MoMA-based BLAS operations outperforming state-of-the-art multi-precision libraries, and MoMA-based NTTs achieving near-ASIC performance on commodity GPUs. By making large integer arithmetic efficient, our work may enable critical innovations in cryptographic algorithms.
\end{enumerate}

\section{Background}

In this section, we begin by outlining the mathematical background of modular and multi-digit arithmetic. Next, we introduce polynomial operations built on modular arithmetic, which represent the primary computational bottlenecks in cryptographic applications like FHE and ZKPs.

\subsection{Modular Arithmetic}
\label{sec:mod_arith}

When operating on an integer ring modulo $q$, $\mathbb{Z}_q$, additions, subtractions, and multiplications of two integers are defined as
\begin{equation}
\begin{aligned}
        c &= a + b  &&\mod q, \\
        c &= a - b  &&\mod q, \\
        c &= ab     &&\mod q,
\end{aligned}
\end{equation}
where $a,b,c\in\mathbb{Z}_q$.
However, modulo operation is significantly more costly than basic operations such as addition and multiplication on standard off-the-shelf hardware. To efficiently implement modular arithmetic, we aim to replace the modulo operations with cheaper operations instead. Using the definition that $0 \leq a < q$ and $0 \leq b < q$ for any $a,b\in\mathbb{Z}_q$, modular addition within the ring can be done by
\begin{equation}
\label{eq:saddmod}
    c =
    \begin{cases}
        a + b - q,      & \text{if } (a + b) > q, \\[.8mm]
        a + b,          & \text{otherwise.}
    \end{cases}
\end{equation}
Modular subtraction can be done by 
\begin{equation}
\label{eq:ssubmod}
    c =
    \begin{cases}
        a - b + q,      & \text{if } a < b, \\[.8mm]
        a - b,          & \text{otherwise.}
    \end{cases}
\end{equation}
For modular multiplication, we use Barrett reduction~\cite{barrett1986implementing}, a popular approach widely used within the cryptography community~\cite{al2022openfhe,boemer2021intel}, for general modulo (that is, not a specific modulus such as Goldilock prime~\cite{hamburg2015ed448}). Modular multiplication using Barrett reduction with the floor operation is defined as 
\begin{equation}
\label{eq:barrett}
    c = ab - \lfloor{ab}/{q}\rfloor q.
\end{equation}
Note that the result is exact. We discuss how to efficiently implement Equation~\ref{eq:barrett} in Section~\ref{sec:soma}.

\subsection{Multi-Digit Arithmetic}
\label{sec:multi_digit}

Multi-digit arithmetic involves the execution of basic mathematical operations (i.e., addition, subtraction, multiplication, and division) on numbers with multiple digits. 
Formally, we define a function $[]_z: \mathbb{Z}^n \to \mathbb{Z}$, parameterized by base digit $z$, as 
\begin{equation}
\label{eq:def_mp}
    [x_0, x_1, \ldots, x_{n-1}]_z = x_0 z^{n-1} + x_1 z^{n-2} + \ldots + x_{n-1} = x.
\end{equation}
For example, in the decimal system where $z = 10$, we can write $[8, 9]_{10} = 8 \cdot 10 + 9 = 89$. We can calculate $n$, given $z$ and $x$, as $n = \lceil\log_z x\rceil$. It is important to note that, at this stage, we are discussing the mathematical concepts without any reference to implementation. This means that the value of $z$ does not necessarily need to fit within a machine word and can be arbitrarily large or small. In Section~\ref{sec:moma}, we will discuss how to efficiently implement multi-digit arithmetic, and how to combine it with modular arithmetic, when each digit is a machine word.

Here, we demonstrate multi-digit arithmetic on 2 digits. Let $a = [a_0, a_1]_z = a_0 z + a_1$ and $b = [b_0, b_1]_z = b_0 z + b_1$. 
The schoolbook multi-digit addition, $c = a + b$, can be written as
\begin{equation}
\begin{aligned}
    &[\delta, c_2]_z = a_1 + b_1, \\
    &[c_0, c_1]_z = a_0 + b_0 + \delta,
\end{aligned}
\end{equation}
where $c = [c_0, c_1, c_2]_z$ and $\delta\in\{0,1\}$.

The schoolbook multi-digit subtraction, $c = a - b$, can be written as
\begin{equation}
\begin{aligned}
    &c_1 = a_1 - b_1, \\
    &\delta = 
    \begin{cases}
        1,          &\text{if } a_1 < b_1, \\[.8mm]
        0,          &\text{otherwise,}    \\
    \end{cases} \\
    &c_0 = a_0 - b_0 - \delta,
\end{aligned}
\end{equation}
where $c = [c_0, c_1]_z$. 

The schoolbook multiplication, $c = ab$, can be written as 
\begin{equation}
\label{eq:muls}
    c = (a_0 b_0)z^2 + (a_0 b_1 + a_1 b_0)z + a_1 b_1,
\end{equation}
where each addition can be further broken down using the aforementioned multi-digit addition. 
The Karatsuba algorithm~\cite{karatsuba1962multiplication} is a divide-and-conquer method for multiplying large numbers 
by recursively breaking down the multiplication of two $n$-digit numbers into three multiplications of $n/2$-digit numbers, along with some additions and subtractions. Formally,
\begin{equation}
\label{eq:mulk}
    c = (a_0 b_0)z^2 + \big((a_0 + a_1)(b_0 + b_1) - a_0 b_0 - a_1 b_1\big)z + a_1 b_1,
\end{equation}
where each addition and subtraction can be further broken down using the aforementioned multi-digit addition and subtraction. 
In $n$-digit arithmetic, the addition of two $n$-digit integers produces a result with at most $n+1$ digits, while subtraction yields a result with at most $n$ digits, and multiplication can produce up to $2n$ digits.

\subsection{Polynomial Operations and NTT}

Polynomial operations and NTT are fundamental cryptographic kernels in advanced encryption schemes such as FHE and ZKPs. In this section, we provide a brief overview of these concepts. 

\paragraph{Polynomial addition and subtraction.} Polynomial operations, especially polynomial multiplications with coefficients reside in $\mathbb{Z}_q$, are the building blocks of advanced cryptographic schemes such as FHE and ZKPs~\cite{satriawan2023conceptual,ben2014succinct}. 
Let $f$ and $g$ denote two polynomials of degree $n$, where $f = \sum_{i=0}^n a_ix^i$ and $g = \sum_{j=0}^n b_jx^j$. 
The addition and subtraction of $f$ and $g$ are defined as the point-wise addition and subtraction of both polynomials' coefficients ($a_i$ and $b_j$), respectively. Prior work~\cite{al2022openfhe} has shown that point-wise multiplication of both polynomials' coefficients is also commonly utilized in FHE schemes. Each of the above three operations can be efficiently implemented using vector operations, where a vector of length $n+1$ represents the coefficients of the degree $n$ polynomial.

We can then utilize the BLAS abstraction~\cite{blackford2002updated} to describe point-wise polynomial operations. Vector addition and subtraction can be interpreted as variants of a BLAS Level 1 operation known as axpy, which is formally defined as 
\begin{equation}
    y = a x + y,
\end{equation}
where $x$ and $y$ are vectors and $a$ is a scalar. Point-wise vector multiplication can be seen as a special case of gemv, a BLAS Level 2 operation that computes a general matrix-vector multiplication. 

\paragraph{Polynomial multiplication.} The multiplication of $f$ and $g$ is defined as
\begin{equation}
    f(x)g(x) = \sum^{2n}_{j = 0}  \sum^{j}_{i = 0} a_i b_{j-i} x^j,
\end{equation}
which is an $O(n^2)$ operation. 
Similarly to how the Fourier transform converts a signal from the time domain to the frequency domain, NTT transforms a polynomial from its coefficient form (e.g., \( f(x) = x^3 + 5x^2 + 2x + 1 \bmod 3 \)) to its evaluation form (e.g., \( \{f(0), f(1), f(2), f(3), f(4)\} \)), thereby reducing the time complexity of polynomial multiplication from $O(n^2)$ to \( O(n \log n) \).
Formally, an $n$-point NTT is defined as
\begin{equation}
    y(k) = \sum^{n-1}_{j=0} x(j) \omega^{jk}_{n} \!\!\!\! \mod p,\quad \ 0 \leq k \leq n-1,
\label{eq:ntt}
\end{equation}
where $\omega_n$ is the $n$-th primitive root of unity. 
As advanced encryption schemes heavily rely on polynomial arithmetic that comes with a prohibited computational overhead, NTT has been widely adopted to accelerate polynomial multiplications. Prior work has shown that NTT accounts for over 90\% of FHE-based application execution time in practice~\cite{fan2023tensorfhe} and around 30\% of execution time for ZKP-based workloads~\cite{zhang2021pipezk}. 
Therefore, we focus on the previously mentioned BLAS operations and NTT as our primary cryptographic kernels, as they encompass a majority of practical cryptographic workloads~\cite{al2022openfhe}.

\section{Multi-word Modular Arithmetic}
\label{sec:moma}

We define a machine word as the largest integer data type that is efficiently supported by the instruction set or hardware. Typically, a machine word is the largest data type that can fit into a single general-purpose register. For example, on x86-64 architectures, a machine word has 64 bits.
To fully leverage the computing efficiency of natively supported machine words, we propose using these words to construct large integer arithmetic when the integer bit-width exceeds the native machine word width. We combine multi-digit arithmetic with modular arithmetic, treating each machine word as a digit, to develop a system called multi-word modular arithmetic (MoMA).
Using our multi-digit definition from Equation~\ref{eq:def_mp}, in MoMA, we now express an integer $x$ as follows:
\begin{equation}
    x = [x_0, x_1, \ldots, x_{k-1}]_{2^{\omega_0}},
\end{equation}
where $k = \lceil\log_{2^{\omega_0}} x\rceil$ and $\omega_0$ is the machine word width.
For simplicity, we denote this as
\begin{equation}
\label{eq:def_moma}
    x = [x_0^{\omega_0}, x_1^{\omega_0}, \ldots, x_{k-1}^{\omega_0}],
\end{equation}
where each $x_i^{\omega_0}$, for $0 \leq i < k$, is an integer of bit-width $\omega_0$. 
This notation is used to build the formal rule system in Section~\ref{sec:rules}. 

As we map from multi-digit arithmetic to its implementation using machine words, we move from the mathematical level to the algorithm level where overflow must be explicitly managed when intermediate results exceed the machine word width. Addressing these challenges is the primary focus of MoMA and will be thoroughly discussed in this section.
We begin with single-word modular arithmetic, where the input integer $x$ fits within a single machine word. As discussed in Section~\ref{sec:multi_digit}, we have to handle results from addition and multiplication that exceed the machine word width. Next, we discuss double-word modular arithmetic, which builds upon single-word modular arithmetic. Double-word modular arithmetic is fundamental to MoMA, as it allows us to represent an arbitrarily large input integer as a double-word type (with an abstract machine word whose bit-width is half of the input bit-width) and decompose it into single-word operations. MoMA applies this decomposition recursively until the abstract machine word is reduced to an actual machine word that can be executed natively.

\subsection{Single-Word Modular Arithmetic}
\label{sec:soma}

\begin{listing}[t]
\begin{minted}[frame=none, obeytabs=true, tabsize=4, linenos, numbersep=-6pt, escapeinside=||, fontsize=\footnotesize]{c} 
    #define MBITS 60
    typedef uint64_t i64;
    typedef unsigned __int128 i128;

    // addition
    void _sadd(i128 *c, i64 a, i64 b) { 
        *c = (i128) a + (i128) b; }
        
    // modular addition
    void _saddmod(i64 *c, i64 a, i64 b, i64 q) {   
        i128 t; t = (i128) a + (i128) b;
        *c = t > q ? (i64) (t - (i128) q) : (i64) t; }
        
    // subtraction
    void _ssub(i64 *c, i64 a, i64 b) { *c = a - b; }
    
    // modular subtraction
    void _ssubmod(i64 *c, i64 a, i64 b, i64 q) {
        i64 t; t = a - b; *c = a < b ? t + q : t; }
        
    // multiplication
    void _smul(i128 *c, i64 a, i64 b) { 
        *c = (i128) a * (i128) b; }
        
    // modular multiplication using Barrett reduction
    void _smulmod(i64 *c, i64 a, i64 b, i64 q, i64 mu) {
        i128 t, r; t = (i128) a * (i128) b; r = t;
        r >>= (MBITS - 2); r *= (i128) mu;
        r >>= (MBITS + 5); t -= (i128) r * (i128) q;
        // correct off-by-one approx. error
        *c = t > q ? (i64) (t - (i128) q) : (i64) t; }
\end{minted}
\vspace{-2mm}
\caption{Single-word modular arithmetic.}
\label{lst:single_word}
\end{listing}

We begin with single-word arithmetic, where all inputs fit entirely within a machine word. When the single-word data type matches the machine word, the single-word arithmetic is usually fully supported by the compiler, allowing us to directly implement the mathematical definitions from Section~\ref{sec:mod_arith}. Note that to fully support single-word arithmetic, the compiler must also provide a double-word representation. This is necessary to store results from single-word operations, even though full double-word arithmetic (which requires quad-word representation) is typically unavailable. For example, in C and CUDA, operations on \verb|uint64_t| are fully supported, as the 128-bit \verb|unsigned __int128| type exists for overflow handling. 
Utilizing the compiler-supported double-word data type allows compilers (e.g., \verb|nvcc|) to leverage specialized instructions such as add-with-carry for carry propagation during compilation.
In Listing~\ref{lst:single_word}, we illustrate both modular and non-modular arithmetic, using \verb|uint64_t| in C as the single-word data type.

Single-word addition, subtraction, and multiplication are natively supported by the compiler. For modular addition and modular multiplication, we need to cast the input using a double-word datatype because the results might exceed the bit-width of a single word. For modular addition, we implement the arithmetic according to Equation~\ref{eq:saddmod} and for modular subtraction, it is a direct translation from Equation~\ref{eq:ssubmod}. We show how to implement efficient modular multiplication using Barrett reduction according to Equation~\ref{eq:barrett}. 
Note that to implement what is shown in Equation~\ref{eq:barrett} efficiently, we need to carefully choose the implementation of $\lfloor ab/q \rfloor$. The straightforward approach is to use integer division, which truncates the results toward zero in C. However, division is computationally expensive and can be vulnerable to timing attacks if it is not implemented as a constant-time operation on certain architectures.
As a result, the cryptography community has developed methods to implement $\lfloor ab/q \rfloor$ using only multiplication and bit shifts, which are significantly more efficient than division. This approach relies on a precomputed value, $\mu$, derived as follows.
We want to approximate $1/q$ in $\lfloor ab/q \rfloor$ by
\begin{equation}  
    1/q = \mu/2^k,
\end{equation}
so that we can compute $\lfloor ab/q \rfloor$ with $\lfloor ab\mu/2^k \rfloor$ using the right shifts. Therefore, we precompute $\mu$ using
\begin{equation}  
\label{eq:mu}
    \mu = \lfloor 2^k/q \rfloor,
\end{equation}
so that $\mu$ is an integer. Now,
\begin{equation}  
    \mu/2^k = \lfloor 2^k/q \rfloor / 2^k \leq 1/q.
\end{equation}
We need to correct this off-by-one error using a conditional subtraction that corresponds to the last line of code in Listing~\ref{lst:single_word}. 
The entire Barrett reduction now becomes
\begin{equation}    
    ab - \lfloor ab \lfloor 2^k/q \rfloor / 2^k \rfloor q.
\end{equation}
We only need multiplications and shifts to compute Equation~\ref{eq:barrett} rather than using divisions if we precompute $\lfloor 2^k/q \rfloor$ as $\mu$ using division for once for the same modulus $q$. 

\subsection{Double-Word Modular Arithmetic}

\begin{listing}[!htb]
\begin{minted}[frame=none, obeytabs=true, tabsize=4, linenos, numbersep=-6pt, escapeinside=||, fontsize=\footnotesize]{c} 
    // addition: quad = double + double
    void _dadd(i64 *c0, i64 *c1, i64 *c2, i64 *c3, 
               i64 a0, i64 a1, i64 b0, i64 b1) {
        i128 s; int cr; s = (i128) a1 + (i128) b1; 
        *c3 = (i64) s; cr = s >> 64; 
        s = (i128) a0 + (i128) b0 + (i128) cr;
        *c2 = (i64) s; *c1 = s >> 64; *c0 = 0; }
    
    // subtraction
    void _dsub(i64 *c0, i64 *c1, i64 a0, i64 a1, 
               i64 b0, i64 b1) {
        int br; *c1 = a1 - b1; br = a1 < b1; 
        *c0 = a0 - b0 - br; }
    
    // less than
    void _dlt(int *c, i64 a0, i64 a1, i64 b0, i64 b1) {
        int i0, i1, i2, i3; i0 = (a0 < b0); 
        i1 = (a0 == b0); i2 = (a1 < b1); 
        i3 = i1 && i2; *c = i0 |\textbar\textbar| i3; }
    
    // modular addition
    void _daddmod(i64 *c0, i64 *c1, i64 a0, i64 a1, 
                  i64 b0, i64 b1, i64 q0, i64 q1) {
        i64 t0, t1, t2, t3, t4, t5; int i;
        _dadd(&t0, &t1, &t2, &t3, a0, a1, b0, b1); 
        _dlt(&i, q0, q1, t2, t3); 
        _dsub(&t4, &t5, t2, t3, q0, q1); 
        *c0 = i ? t4 : t2; *c1 = i ? t5 : t3; }
    
    // modular subtraction
    void _dsubmod(i64 *c0, i64 *c1, i64 a0, i64 a1, 
                  i64 b0, i64 b1, i64 q0, i64 q1) {
        i64 t0, t1, t2, t3, t4, t5; int i;
        _dsub(&t0, &t1, a0, a1, b0, b1); 
        _dadd(&t2, &t3, &t4, &t5, t0, t1, q0, q1);
        _dlt(&i, a0, a1, b0, b1); 
        *c0 = i ? t4 : t0; *c1 = i ? t5 : t1; }
\end{minted}
\vspace{-2mm}
\caption{Double-word modular addition and subtraction.}
\label{lst:dadd_dsub}
\end{listing}

We define double-word integers as integers with a bit-width of $2\omega$, where $\omega$ denotes the bit-width of a single word. While single-word modular arithmetic is relatively straightforward, as illustrated in Listing~\ref{lst:single_word}, double-word arithmetic is significantly more complex. This complexity arises because adding or multiplying two double-word integers can lead to quad-word (integers with a bit-width of $4\omega$) results, which further complicates arithmetic operations.
We will start with double-word modular addition and subtraction. 

\paragraph{Addition and subtraction.} Listing~\ref{lst:dadd_dsub} shows the C implementation of double-word modular addition and subtraction. As a quad-word cannot be natively represented, we break it down to four single words $a = [a^{64}_0, a^{64}_1, a^{64}_2, a^{64}_3]$ as defined in Equation~\ref{eq:def_moma}. 
In the implementation of \verb|_dadd|, carry extraction and propagation must be handled explicitly in the code, as this cannot be managed automatically by the compiler. Similarly, \verb|_dsub| requires explicit management of the borrow. For the modulo operation, the only missing part is the comparison in the conditional assignments within \verb|_saddmod| and \verb|_ssubmod|. Thus, we implement \verb|_dlt| to perform comparisons between two double words. 
In this specific example, we assume that the single-word data type corresponds to the machine word (i.e., \verb|uint64_t|). This implies that for \verb|_dsub| and \verb|_dlt|, native operations can be utilized if we combine $[a^{64}_0, a^{64}_1]$ and $[b^{64}_0, b^{64}_1]$ into a 128-bit representation and use the natively supported 128-bit subtraction and comparison. However, we present an implementation that does not rely on double-word support to demonstrate how operations are handled when the abstract single word is not the machine word. In Section~\ref{sec:rules}, we can then observe a clear mapping from these double-word operations to formal rewrite rules within MoMA. 

\begin{listing}[!htb]
\begin{minted}[frame=none, obeytabs=true, tabsize=4, linenos, numbersep=-6pt, escapeinside=||, fontsize=\footnotesize]{c} 
    // addition: quad = quad + quad
    void _qadd(i64 *c0, i64 *c1, i64 *c2, i64 *c3, 
               i64 a0, i64 a1, i64 a2, i64 a3,
               i64 b0, i64 b1, i64 b2, i64 b3) {
        i128 s; i64 t0, t1; int cr;
        _dadd(&t0, &t1, c2, c3, a2, a3, b2, b3);
        // t1 is either 0 or 1
        s = (i128) a1 + (i128) b1 + (i128) t1;
        *c1 = (i64) s; cr = s >> 64;
        s = (i128) a0 + (i128) b0 + (i128) cr; 
        *c0 = (i64) s; }
    
    // schoolbook multiplication
    void _dmuls(i64 *c0, i64 *c1, i64 *c2, i64 *c3, 
                i64 a0, i64 a1, i64 b0, i64 b1) {
        i64 t0, t1, t2, t3, t4, t5, t6, t7, 
            t8, t9, t10, t11; i128 s; 
        s = (i128) a1 * (i128) b1; 
        t0 = s >> 64; t1 = (i64) s;
        s = (i128) a0 * (i128) b0; 
        t2 = s >> 64; t3 = (i64) s;
        s = (i128) a0 * (i128) b1; 
        t4 = s >> 64; t5 = (i64) s;
        s = (i128) a1 * (i128) b0; 
        t6 = s >> 64; t7 = (i64) s;
        // a0b1 + a1b0
        _dadd(&t8, &t9, &t10, &t11, t4, t5, t6, t7); 
        // a0b0z^2 + (a0b1 + a1b0)z + a1b1
        _qadd(c0, c1, c2, c3, t2, t3, t0, t1, 
              t9, t10, t11, 0); }
\end{minted}
\vspace{-2mm}
\caption{Double-word schoolbook multiplication.}
\label{lst:dmuls}
\end{listing}

\paragraph{Schoolbook multiplication.} 
In Listing~\ref{lst:dmuls}, we present the implementation of schoolbook multiplication, denoted as \verb|_dmuls|, as defined by Equation~\ref{eq:muls}. In MoMA, representing large integers as multiple words simplifies operations such as shifting by one or multiples of word width. For example, shifting $a = [a^{64}_0, a^{64}_1, a^{64}_2, a^{64}_3]$ left by one word width results in $[a^{64}_1, a^{64}_2, a^{64}_3, 0]$. To complete Equation~\ref{eq:muls}, we need to support the addition of two quad-words. This is achieved through \verb|_qadd|, which extends \verb|_dadd| by applying the same multi-word addition strategy. 
We have omitted the example code for Karatsuba multiplication due to space constraints; however, it can be derived using Equation~\ref{eq:mulk}.

\begin{listing}[!htb]
\begin{minted}[frame=none, obeytabs=true, tabsize=4, linenos, numbersep=-6pt, escapeinside=||, fontsize=\footnotesize]{c} 
    #define MBITS 124
    
    // right shift: double = quad >> k, where k in [64, 128]
    void _qshr(i64 *c0, i64 *c1, i64 a0, i64 a1, 
               i64 a2, i64 a3, int b) {
        i64 t0, t1, t2, t2, t4, t5, t6, t7; int i0, i1;
        i0 = b - 64; i1 = 128 - b; t0 = a2 >> i0; 
        /* a mask of 1s */
        t1 = (i64) 1; t2 = t1 << i0; t3 = t2 - 1; 
        t4 = a0 & t3; t5 = t4 << i1; t6 = a1 >> i0;
        *c0 = t5 |\textbar| t6; t7 = a1 << i1; *c1 = t7 |\textbar\textbar| t0; }
    
    // modular multiplication using Barrett reduction
    void _dmulmod(i64 *c0, i64 *c1, i64 a0, i64 a1, 
                  i64 b0, i64 b1, i64 q0, i64 q1, 
                  i64 mu0, i64 mu1) {
        i64 t0, t1, t2, t3, t4, t5, t6, t7, 
            t8, t9, t10, t11, t12, t13, t14, t15, 
            t16, t17, t18, t19, t20, t21; int i;
        _dmuls(&t0, &t1, &t2, &t3, a0, a1, b0, b1);
        _qshr(&t4, &t5, t0, t1, t2, t3, MBITS-2)
        // t8, t9 will not be used
        _dmuls(&t6, &t7, &t8, &t9, t4, t5, mu0, mu1);
        // [t10, t13] = [t6, t7, t8, t9] >> MBITS+5
        t10 = t6 >> 1; t11 = t6 << 63; 
        t12 = t7 >> 1; t13 = t11 |\textbar| t12;
        // t14, t15 will not be used
        _dmuls(&t14, &t15, &t16, &t17, t10, t13, q0, q1);
        // optimization given that the first half matches
        _dsub(&t18, &t19, t2, t3, t16, t17);
        _dsub(&t20, &t21, t18, t19, q0, q1);
        _dlt(&i, t18, t19, q0, q1)
        *c0 = i ? t18 : t20; *c1 = i ? t19 : t21; }
\end{minted}
\vspace{-2mm}
\caption{Double-word modular multiplication.}
\label{lst:dmulmod}
\end{listing}

\paragraph{Modular multiplication.} The most complex operation for double-word arithmetic is modular multiplication. As illustrated in \verb|_dmulmod| in Listing~\ref{lst:dmulmod}, it involves three multiplications, two right shifts by non-multiples of the word width, and a final conditional subtraction. 
\verb|_dmulmod| is built upon \verb|_smulmod| from Listing~\ref{lst:single_word}. To handle shifting by $\texttt{MBITS} - 2$, we use \verb|_qshr|, which shifts a quad-word by $k$ bits, where $k$ ranges from one single-word width to two single-word width.  Shifting right by $\texttt{MBITS} + 5$ is more straightforward, as it involves discarding the lower part of the quad-word. Note that in the second multiplication, the lower part of the result is discarded due to the subsequent shift operation.

To implement the double-word version of \begin{small}\verb|t -= r * q|\end{small} in \verb|_smulmod|, we need to perform a subtraction between two quad-words. According to Barrett reduction, $t$ can only be either $c$ or $c + q$, where $c = ab \bmod q$. Since $c$ is less than $q$ and $q$ has a bit-width smaller than the word width, both $c$ and $c + q$ are less than the word width. Consequently, we only need to subtract the lower part of $rq$ from the lower part of $t$, because the final result will fit within a double word. Therefore, \verb|_dsub| is used, and the higher part of the third multiplication is not needed.

\paragraph{Multi-word modular arithmetic via recursion.} 
Given our formal definition of double-word modular arithmetic, we can now define MoMA through recursion. Let the bit-width of the input integer be $\lambda$. We start by applying double-word modular arithmetic to break it down into equivalent computations using data types with bit-width $\lambda/2$. This process is recursively applied to the resulting $\lambda/2$ bit-width data types, continuing until $(\lambda/2^k) \leq \omega_0$, where $\omega_0$ is the machine word width and $k$ is the number of recursion steps. For example, if the input integer $a$ is 512 bits and the machine word width is 64 bits, three recursion steps are required. The data type breakdown would proceed as follows:
$$
    a = [a^{256}_0, a_1^{256}] = [a^{128}_0, a_1^{128}, a^{128}_2, a_3^{128}] = [a^{64}_0, a_1^{64}, \ldots, a_7^{64}].
$$
The complexity of the associated computations will increase significantly as we recursively break down the data type.

\begin{table*}[t]
    \caption{Multi-word modular arithmetic core rewrite rules.}
    \vspace{-1mm}
    \label{tab:rules}
    \centering
    \begin{minipage}{0.86\textwidth}
    \rule{\textwidth}{.5pt}
    \begin{align}
        \begin{split}
        \label{eq:r1}
            a^{2\omega} \rewriteto & [a^\omega_0, a_1^\omega]
        \end{split}
        \\[1mm]
        \begin{split}
        \label{eq:exhigh}
            c_0^\omega = \lfloor[a_0^\omega, a_1^\omega]/2^\omega\rfloor \rewriteto & c_0^\omega = a_0^\omega
        \end{split}
        \\[1mm]
        \begin{split}
        \label{eq:exlow}
            c_0^\omega = [a_0^\omega, a_1^\omega]\omod 2^\omega \rewriteto & c_0^\omega = a_1^\omega
        \end{split}
        \\[1mm]
        \begin{split}
        \label{eq:addqdd}
            [c_0^1, c_1^\omega, c_2^\omega] = [a^\omega_0, a_1^\omega] + [b^\omega_0, b_1^\omega] \rewriteto & [\delta_0^1, c_2^\omega] = a_1^\omega + b_1^\omega,\; [c^1_0, c_1^\omega] = \delta_0^1 + a^\omega_0 + b_0^\omega
        \end{split}
        \\[1mm]
        \begin{split}
        \label{eq:adc}
            [c_0^1, c_1^\omega] = a_1^\omega + b_1^\omega \rewriteto & c_0^1 = \lfloor(a_1^\omega + b_1^\omega)/2^\omega\rfloor,\; c_1^\omega = (a_1^\omega + b_1^\omega)\omod 2^\omega
        \end{split}
        \\[1mm]
        \begin{split}
        \label{eq:moddq_add}
            [c^\omega_0, c^\omega_1] = [a_0^1, a_1^\omega, a_2^\omega] \omod [q^\omega_0, q_1^\omega] \rewriteto & \delta^1_0 = [q^\omega_0, q_1^\omega] < [a_1^\omega, a_2^\omega], \\
            & \delta^1_1 = (0 < a_0^1) \vee \big((a_0^1 \qeq 0) \wedge \delta_0^1\big), \\ 
            & [b_0^\omega, b_1^\omega] = [a_1^\omega, a_2^\omega] - [q^\omega_0, q_1^\omega], \\
            & [c_0^\omega, c_1^\omega] =
            \begin{cases}
                [b_0^\omega, b_1^\omega], \quad \text{if } \delta^1_1 \qeq 1, \\[.8mm]
                [a_1^\omega, a_2^\omega], \quad \text{otherwise}
            \end{cases}
        \end{split}
        \\[1mm]
        \begin{split}
        \label{eq:subddd}
            [c^\omega_0, c^\omega_1] = [a_0^\omega, a_1^\omega] - [b_0^\omega, b_1^\omega] \rewriteto & c_1^\omega = a_1^\omega - b_1^\omega,\; \delta_0^1 = a_1^\omega < b_1^\omega,\; c_0^\omega = a_0^\omega - b_0^\omega - \delta_0^1
        \end{split}
        \\[1mm]
        \begin{split}
        \label{eq:ltd}
            \delta_0^1 = [a_0^\omega, a_1^\omega] < [b_0^\omega, b_1^\omega] \rewriteto & \delta_0^1 = (a_0^\omega < b_0^\omega) \vee \big((a_0^\omega \qeq b_0^\omega) \wedge (a_1^\omega < b_1^\omega)\big)
        \end{split}
        \\[1mm]
        \begin{split}
        \label{eq:eqd}
            \delta_0^1 = [a_0^\omega, a_1^\omega] \qeq [b_0^\omega, b_1^\omega] \rewriteto & (a_0^\omega \qeq b_0^\omega) \wedge (a_1^\omega \qeq b_1^\omega)
        \end{split}
        \\[1mm]
        \begin{split}
        \label{eq:mulqdd}
            [c_0^\omega, c_1^\omega, c_2^\omega, c_3^\omega] = [a_0^\omega, a_1^\omega] \cdot [b_0^\omega, b_1^\omega] \rewriteto & [d_0^\omega, d_1^\omega] = a_1^\omega \cdot b_1^\omega,\; [e_0^\omega, e_1^\omega] = a_0^\omega \cdot b_0^\omega, \\
            & [f_0^\omega, f_1^\omega] = a_0^\omega \cdot b_1^\omega,\; [g_0^\omega, g_1^\omega] = a_1^\omega \cdot b_0^\omega, \\
            & [h_0^1, h_1^\omega, h_2^\omega] = [f_0^\omega, f_1^\omega] + [g_0^\omega, g_1^\omega], \\
            & [c_0^\omega, c_1^\omega, c_2^\omega, c_3^\omega] = [e_0^\omega, e_1^\omega, d_0^\omega, d_1^\omega] + [h_0^1, h_1^\omega, h_2^\omega, 0]
        \end{split}
        \\[1mm]
        \begin{split}
        \label{eq:addqqq}
            [c_0^\omega, c_1^\omega, c_2^\omega, c_3^\omega] = [a_{0{\text -}3}^\omega] + [b_{0{\text -}3}^\omega] \rewriteto & [\delta_0^1, c_3^\omega] = a_3^\omega + b_3^\omega,\; [\delta_1^1, c_2^\omega] = a_2^\omega + b_2^\omega + \delta_0^1, \\
            & [\delta_2^1, c_1^\omega] = a_1^\omega + b_1^\omega + \delta_1^1,\; [0, c_0^\omega] = a_0^\omega + b_0^\omega + \delta_2^1
        \end{split}
    \end{align}
    \rule{\textwidth}{.5pt}
    We use the symbol of equality ($=$) to denote assignment and the symbol of equality with a question mark ($=_{\scaleto{?}{4pt}}$) to indicate a comparison of equality. We assume that any comparison evaluates to $1$ if true and $0$ if false. For brevity, we use $[a_{0{\text -}n}^\omega]$ to represent the sequence $[a_0^\omega, \ldots, a_n^\omega]$.
    \end{minipage}
    \vspace{-2mm}
\end{table*}

\section{Code Generation: Rewriting on Data Types}
\label{sec:rules}

To implement MoMA, we introduce a rule system composed of numerous rewrite rules that operate on integer data types. This system recursively decomposes operations involving large data types into equivalent operations using smaller data types until all operations are conducted with data types natively supported by the machine. In this section, we formally define the rule system, which can be integrated into compiler infrastructures and code generators. At each recursive step, we treat the current maximal integer data type as a double word, as discussed in Section~\ref{sec:moma}, and decompose the computations into equivalent operations with single words. This process continues until the single word at that step matches the machine word width. Therefore, to apply the rule system as a program transformation pass, we assume that the input bit-width for application kernels is known at compile/code generation time.

In each recursion step, let the single word width be denoted by $\omega$ and the double word width is $2\omega$. 
We use $x^{\omega}$ to represent a single-word integer of bit-width $\omega$. 
For instance, when $\omega = 256$, a double-word integer can be defined as $a^{512}$ and decomposed into two single-word integers using the definition from Equation~\ref{eq:def_moma}: $a^{512} = [a_0^{256}, a_1^{256}]$.
This data type breakdown process is formally defined as rewrite rule (\ref{eq:r1}). For more complex rules, we use the symbol of equality ($=$) to denote assignment. For example, $c^{2\omega} = a^\omega + b^\omega$ indicates that the result of $a^\omega + b^\omega$ is assigned to $c^{2\omega}$. 
To denote the comparison of equality, we use the equality symbol with a question mark ($=_{\scaleto{?}{4pt}}$). We use $\delta^1$ to represent the result of a comparison or a carry/borrow bit, with the assumption that any comparison evaluates to $1$ when true and $0$ when false. 

The core rewrite rules used to implement MoMA are presented in Table~\ref{tab:rules}. 
Since each rule reduces the integer bit-width required during computation, the largest data type incurred from the right-hand side computations is always smaller than the largest data type incurred from the left-hand side operation. On the right-hand side, the sequence of computations is crucial, as each operation must be executed from left to right and top to bottom.

\paragraph{Example: rewriting modular addition.}
We now show a concrete example of applying the rewrite rules shown in Table~\ref{tab:rules} to break down a double-word modular addition:
\begin{equation}
\label{eq:daddmod}
    c^{2\omega} = (a^{2\omega} + b^{2\omega}) \omod q^{2\omega},
\end{equation}
where $q$ is the modulus. When $\omega = 64$, we can implement $a^\omega$ using the \verb|uint64_t| data type and $a^{2\omega}$ \sloppy{using the \verb|unsigned __int128| data type} in C. The above operation can then be directly mapped to the implementation of \verb|_daddmod| in Listing~\ref{lst:dadd_dsub}:
\begin{minted}[frame=none, obeytabs=true, tabsize=4, linenos, numbersep=-6pt, escapeinside=||, fontsize=\small]{c} 
    void _daddmod(i64 *c0, i64 *c1, i64 a0, i64 a1, 
                  i64 b0, i64 b1, i64 q0, i64 q1)
\end{minted}
For illustration purposes, in this example we consider $\omega$ to be the bit-width of the abstract single word at a recursion step that is not the final recursion step, that is, $\omega > \omega_0$, where $\omega_0$ is the machine word width.

To begin with, by (\ref{eq:r1}), we can decompose $c^{2\omega}$ into $[c^\omega_0, c^\omega_1]$ by performing floor division and modulo operation with $2^\omega$ to extract the higher and lower parts, respectively. These operations are formally represented by (\ref{eq:exhigh}) and (\ref{eq:exlow}). 
Note that, when dealing with an abstract single word during intermediate recursion steps, we do not need to explicitly perform floor division and modulo operations. Instead, we can conceptually transform a single variable $c^{2\omega}$ into an array of two variables, $[c^\omega_0, c^\omega_1]$. It is only at the final recursion step, when the abstract representation must be concretized into actual code, that we need to implement these operations explicitly. Specifically, we use a right shift by $\omega_0$ to extract the higher part and typecasting to $\mathsf{T}_{\omega_0}$ to extract the lower part, where $\mathsf{T}_{\omega_0}$ represents the integer data type corresponding to the machine word width $\omega_0$. For instance, in C, if $\omega = 64$, then $\mathsf{T}_{\omega_0}$ is \verb|uint64_t|. 
We also apply (\ref{eq:r1}) to $a^{2\omega}, b^{2\omega}, q^{2\omega}$ and obtain
\begin{equation}
    [c^\omega_0, c^\omega_1] = ([a^\omega_0, a_1^\omega] + [b^\omega_0, b_1^\omega]) \omod [q^\omega_0, q_1^\omega].
\end{equation}
Then, to compute a double-word addition $[a^\omega_0, a_1^\omega] + [b^\omega_0, b_1^\omega]$, we apply (\ref{eq:addqdd}) which in turn uses (\ref{eq:adc}) to break down double-word addition to single-word additions. 
Thus, we have rewrited (\ref{eq:daddmod}) into 
\begin{equation}
\begin{aligned}
    [\delta_0^1, d_2^\omega] &= a_1^\omega + b_1^\omega, \\
    [d^1_0, d_1^\omega] &= \delta_0^1 + a^\omega_0 + b_0^\omega, \\
    [c^\omega_0, c^\omega_1] &= [d_0^1, d_1^\omega, d_2^\omega] \omod [q^\omega_0, q_1^\omega].
\end{aligned}
\end{equation}
We then need to rewrite $[d_0^1, d_1^\omega, d_2^\omega] \omod [q^\omega_0, q_1^\omega]$ as native modulo support is not available for an integer with a bit-width of $2\omega+1$. As we introduced earlier in Section~\ref{sec:mod_arith}, modulo after addition can be computed by a comparison, a subtraction, and a conditional assignment. Therefore, by (\ref{eq:moddq_add}), we have
\begin{equation}
\begin{aligned}
    [\delta_0^1, d_2^\omega] &= a_1^\omega + b_1^\omega, \\
    [d^1_0, d_1^\omega] &= \delta_0^1 + a^\omega_0 + b_0^\omega, \\
    \delta^1_0 &= [q^\omega_0, q_1^\omega] < [d_1^\omega, d_2^\omega], \\
    \delta^1_1 &= (0 < d_0^1) \vee \big((d_0^1 \qeq 0) \wedge \delta_0^1\big), \\
    [f_0^\omega, f_1^\omega] &= [d_1^\omega, d_2^\omega] - [q^\omega_0, q_1^\omega], \\
    [c_0^\omega, c_1^\omega] &=
    \begin{cases}
        [f_0^\omega, f_1^\omega], \quad \text{if } \delta^1_1 \qeq 1, \\[.8mm]
        [d_1^\omega, d_2^\omega], \quad \text{otherwise.}
    \end{cases}
\end{aligned}
\end{equation}
Lastly, there are two double-word computations that need to be broken down: i) $\delta^1_0 = [q^\omega_0, q_1^\omega] < [d_1^\omega, d_2^\omega]$ and ii) $[f_0^\omega, f_1^\omega] = [d_1^\omega, d_2^\omega] - [q^\omega_0, q_1^\omega]$. 
Using (\ref{eq:subddd}) and (\ref{eq:ltd}), we obtain
\begin{equation}
\label{eq:final_addmod}
\begin{aligned}
    [\delta_0^1, d_2^\omega] &= a_1^\omega + b_1^\omega, \\
    [d^1_0, d_1^\omega] &= \delta_0^1 + a^\omega_0 + b_0^\omega, \\
    \delta_0^1 &= (q_0^\omega < d_1^\omega) \vee \big((q_0^\omega \qeq d_1^\omega) \wedge (q_1^\omega < d_2^\omega)\big), \\
    \delta^1_1 &= (0 < d_0^1) \vee \big((d_0^1 \qeq 0) \wedge \delta_0^1\big), \\
    f_1^\omega &= d_2^\omega - q_2^\omega, \delta_0^1 = d^\omega_2 < q_2^\omega, f_0^\omega = d_1^\omega - q_1^\omega - \delta_0^1, \\
    [c_0^\omega, c_1^\omega] &=
    \begin{cases}
        [f_0^\omega, f_1^\omega], \quad \text{if } \delta^1_1 \qeq 1, \\[.8mm]
        [d_1^\omega, d_2^\omega], \quad \text{otherwise.}
    \end{cases}
\end{aligned}
\end{equation}
Note that double-word assignments, such as \([c_0^\omega, c_1^\omega] = [a_0^\omega, b_1^\omega]\), can be trivially implemented by individually assigning the corresponding single words. The same applies to conditional assignments. Consequently, we do not explicitly list these transformations as rewrite rules.

The above sequence of computations directly maps to \verb|_daddmod| in Listing~\ref{lst:dadd_dsub} which consists of \verb|_add|, \verb|_dlt|, \verb|_dsub| and a conditional assignment at the end.
In other words, when \(\omega = 64\), Equation~\ref{eq:final_addmod} can be implemented exactly as shown for \verb|_daddmod|. However, if \(\omega\) exceeds the machine word width and further recursion is required, an additional rule (\ref{eq:eqd}) is needed to break down the equality comparison in (\ref{eq:moddq_add}) and (\ref{eq:ltd}).

In summary, we use the rules from (\ref{eq:r1}) to (\ref{eq:eqd}) to decompose double-word modular addition. The same set of rules is sufficient to break down double-word modular subtraction. For double-word multiplication (without modulo), two additional rules, (\ref{eq:mulqdd}) and (\ref{eq:addqqq}), are necessary. Due to space constraints, the remaining rules developed for MoMA are omitted.

\paragraph{Non-power-of-two input bit-widths.}
The proposed mathematical formalism directly enables optimizations for inputs with non-power-of-two bit-widths, which is common in ZKP applications where the input bit-widths are often 381 bits (e.g., for the BLS12-381 elliptic curve) or 753 bits (e.g., for the MNT4753 elliptic curve).
Let the non-power-of-two input bit-width be denoted by \(\lambda\), and let \(\omega\) be the closest power-of-two less than \(\lambda\), i.e., \(\omega < \lambda < 2\omega\). In this case, while using the data type \(\mathsf{T}^{2\omega}\) to represent the input with bit-width \(\lambda\) allows us to apply MoMA rewrite rules, many operations will involve zeros and/or evaluate to zero at runtime. Additionally, the number of these redundant operations cascades as the recursion depth required to reduce the bit-width to machine words grows, particularly when \((2\omega - \lambda) > (\lambda - \omega)\). For example, when \(\lambda = 576\) and \(\omega = 512\), 448 bits per input are zero at runtime, offering room for optimization by pruning no-ops during code generation.

Since we assume the input bit-width is known at compile/code generation time, it is straightforward to optimize operations on redundant bits using MoMA and the associated rule system. Let $\omega_0$ denote the machine word width, which is typically a power of two. For any bit-width $\lambda$ such that $\omega < \lambda < 2\omega$, we represent $\lambda$ using $k\omega_0$, where $(k-1)\omega_0 < \lambda < k\omega_0$. Using the multi-word representation defined in Equation~\ref{eq:def_moma} and let $x$ be the input with bit-width $\lambda$, we have
\begin{equation}
    x = [0,\ldots,0,\,x_0^{\omega_0},\ldots,x_{k-1}^{\omega_0}].
\end{equation}
For example, when $\lambda = 753$ and $\omega_0 = 64$, $x$ can be written as
\begin{equation}
    [0,\, 0,\, 0,\, 0,\, x_0^{64},\ldots,x_{11}^{64}],
\end{equation}
or equivalently, $[0,\, 0,\, x_0^{128},\ldots,x_{5}^{128}]$ and $[0,\, x_0^{256},x_1^{256},x_{2}^{256}]$.
This implies that, during the recursive application of MoMA rewrite rules, we can set many single words (with bit-width $\omega$) to zero at certain recursion levels, thereby pruning many operations at compile time. 
We employed this optimization when evaluating NTTs for 384-bit and 768-bit inputs.

\begin{figure*}[t]
\centering
    \subfloat[128-bit]{\includegraphics[width=0.5\columnwidth,trim={5mm 0mm 5mm 5mm},clip]{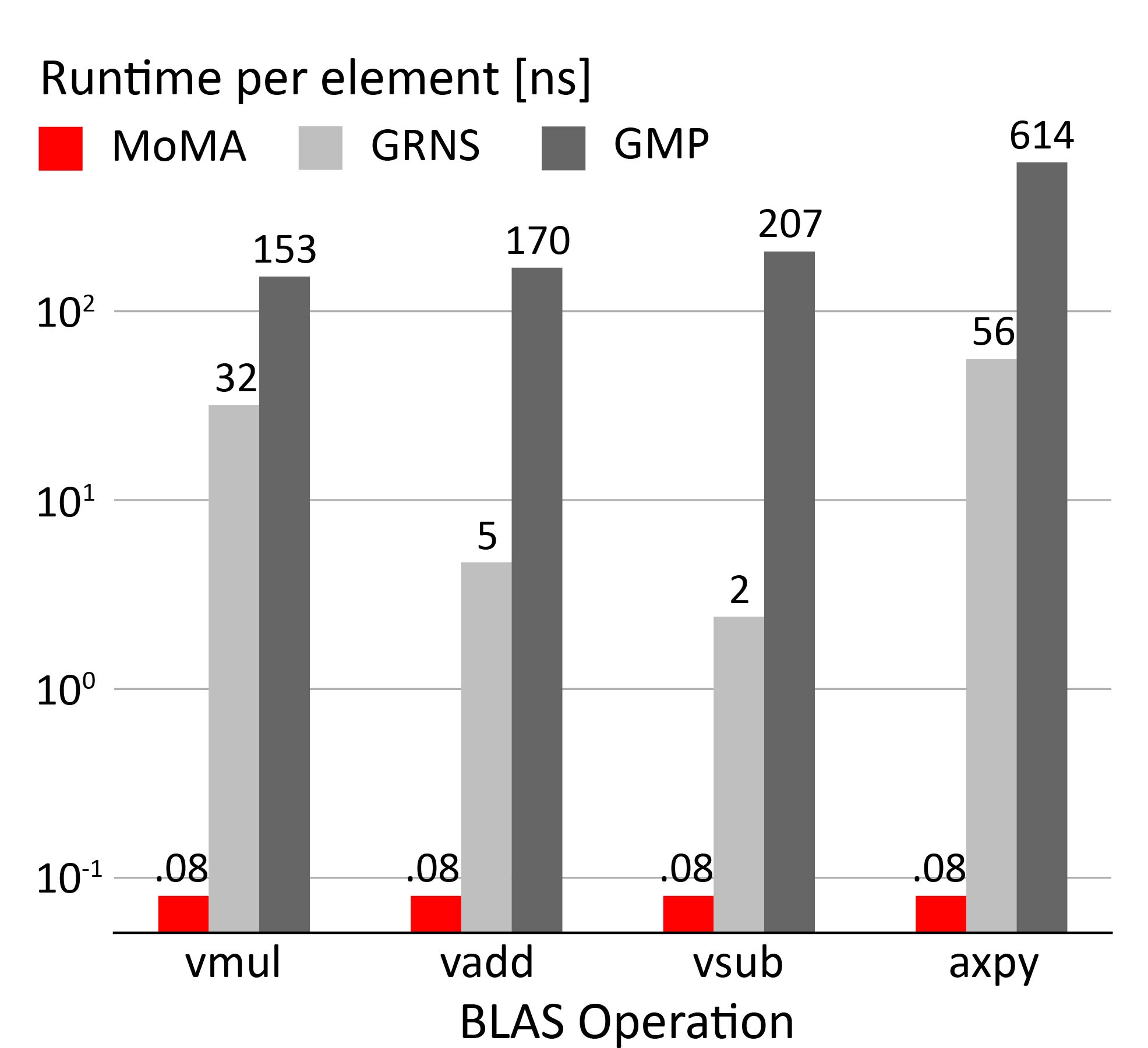}}
    \hspace{0.8mm}
    \subfloat[256-bit]{\includegraphics[width=0.5\columnwidth,trim={5mm 0mm 5mm 5mm},clip]{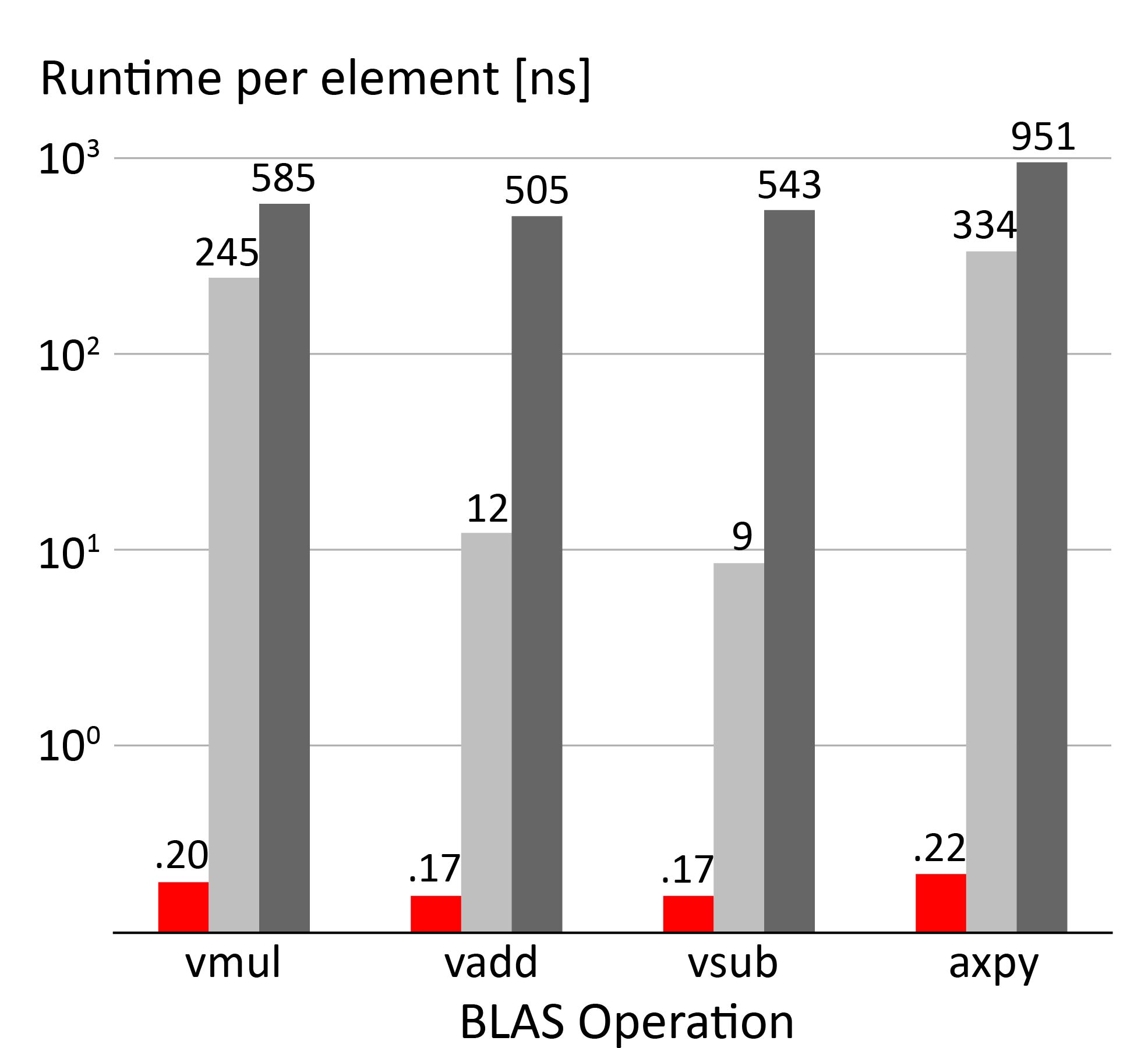}}
    \hspace{0.8mm}
    \subfloat[512-bit]{\includegraphics[width=0.5\columnwidth,trim={5mm 0mm 5mm 5mm},clip]{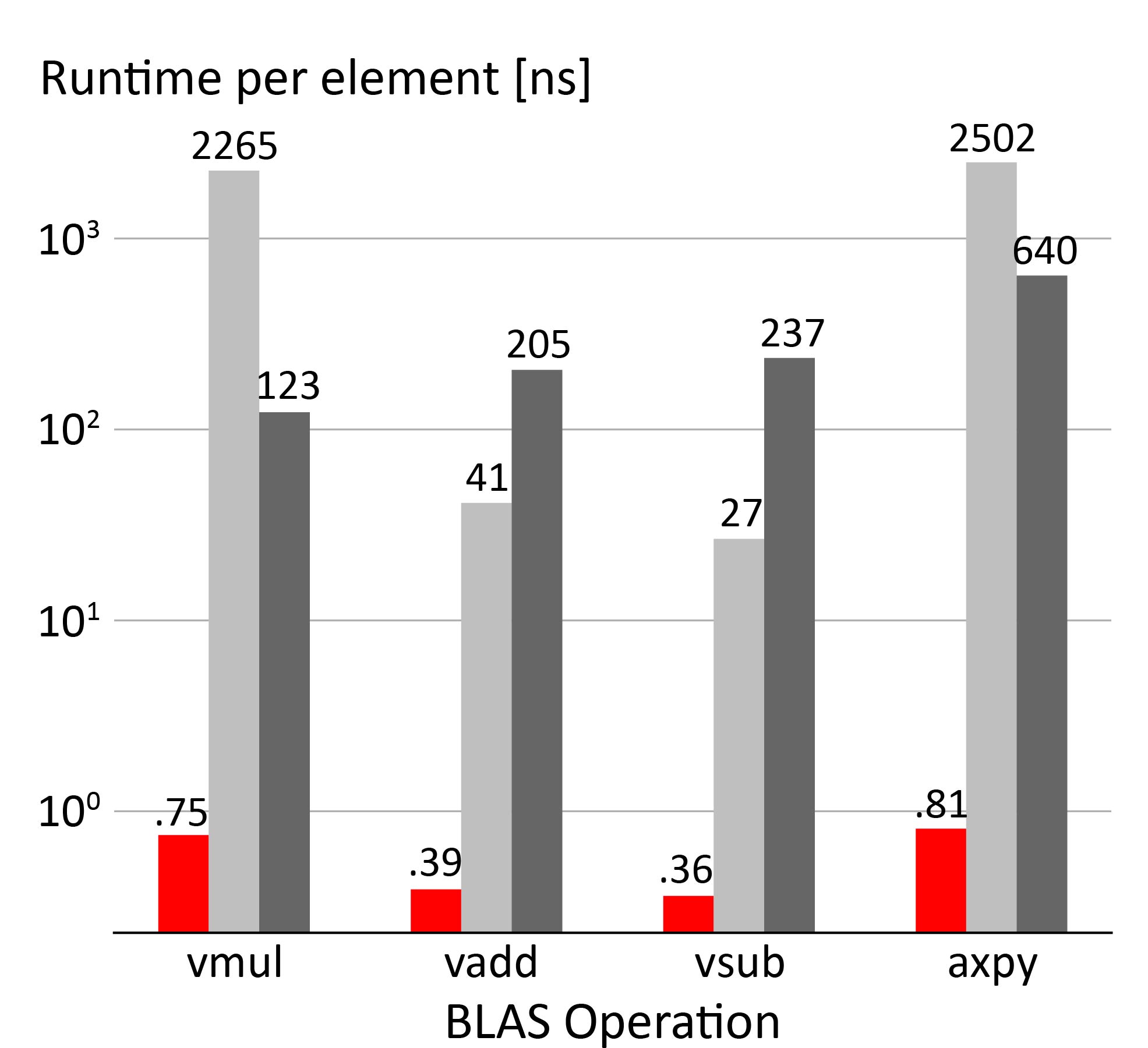}}
    \hspace{0.8mm}
    \subfloat[1,024-bit]{\includegraphics[width=0.5\columnwidth,trim={5mm 0mm 5mm 5mm},clip]{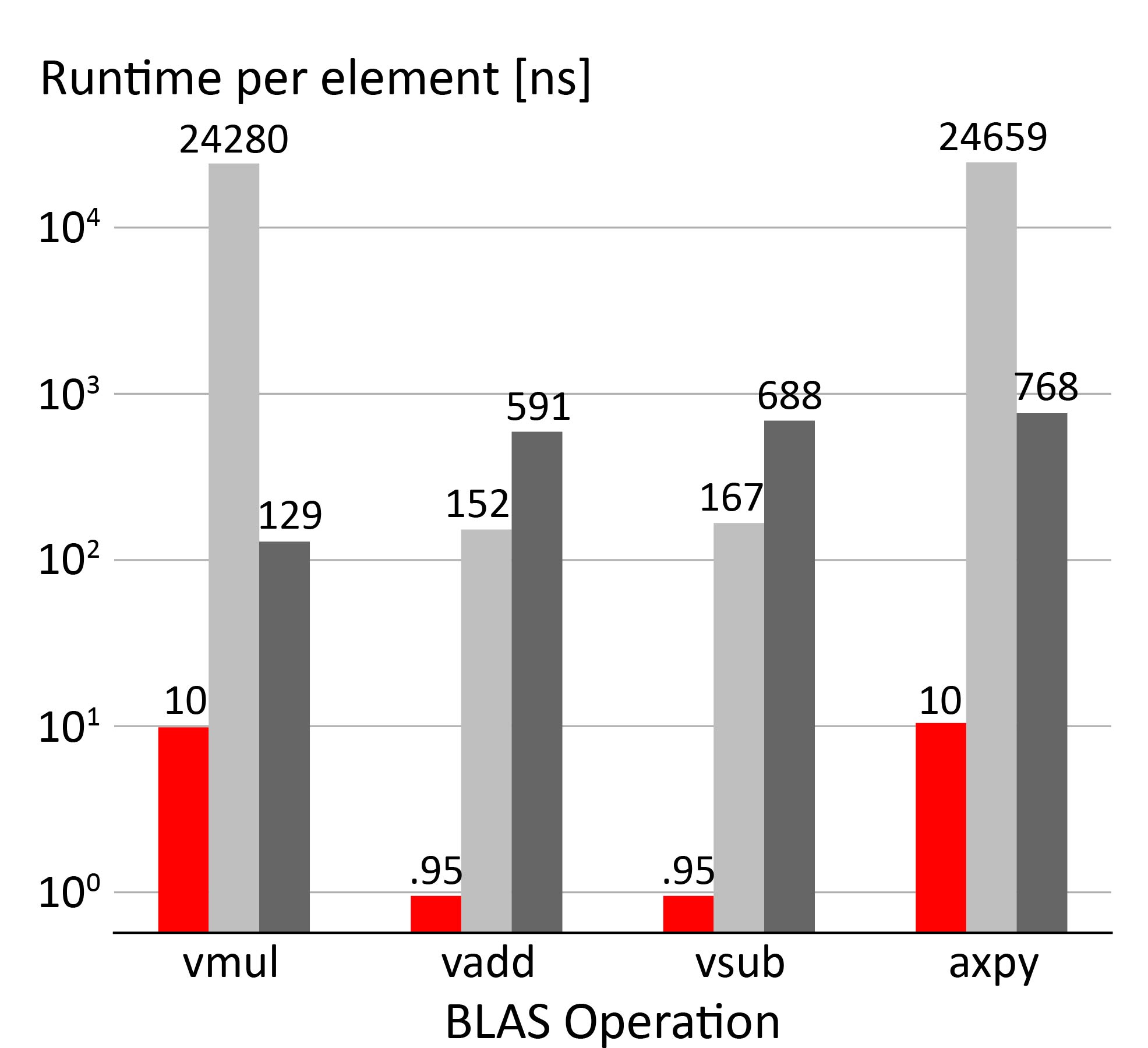}}
    \vspace{-2mm}
    \caption{Performance of BLAS operations with various input bit-widths on CPU and GPU.}
    \label{fig:blas_results}
    \vspace{-3mm}
\end{figure*}

\paragraph{Implementation.} 
We implement MoMA using the rewrite rules detailed in this section through a high-performance code generator, SPIRAL~\cite{franchetti2018spiral}. SPIRAL converts high-level mathematical specifications into highly optimized code for various architectures. It provides a declarative and platform-agnostic mathematical language that offers excellent support for implementing mathematically formal rule systems. 
Specifically, we implement the MoMA rewrite rules as a recursive code generation pass that operates at the abstract code level in SPIRAL. Given the input bit-width and the machine word width, this pass generates equivalent abstract code where each variable's data type is natively supported by the machine. We build on the SPIRAL NTTX package~\cite{zhang2023generating} to map the final abstract code to efficient CUDA implementations on GPUs. 

\section{Evaluation}

In this section, we describe the experimental setup and discuss the results of applying the MoMA rule system to implement BLAS operations and NTT for inputs with large bit-widths. 

\subsection{Experimental Setup}

Prior work~\cite{zhang2023nttongpu} introduced a code generation pass in SPIRAL targeting NVIDIA GPUs for NTTs (encapsulated within the SPIRAL NTTX package~\cite{zhang2023generating}). However, this work was limited to NTTs and only supported input bit-widths twice the machine word width (e.g., 128 bits on NVIDIA GPUs), relying on manually implemented double-word modular arithmetic functions.

We extended this work by implementing the MoMA rule system as a recursive program transformation pass, enabling the NTTX package to handle much larger bit-widths. Additionally, we expanded the SPIRAL NTTX package to generate efficient implementations for BLAS operations, including vector addition, subtraction, multiplication, and axpy (vector-scalar product followed by vector addition). This extension significantly broadens the NTTX package’s capabilities, allowing it to support a wide range of cryptographic workloads through efficient polynomial arithmetic.

\paragraph{Parallelization techniques.}
MoMA, built on top of the SPIRAL NTTX package~\cite{zhang2023generating,zhang2023nttongpu}, leverages the parallelization techniques provided by the framework. For NTTs, each CUDA thread processes one or more butterfly operations in each stage of the NTT, as there are no data dependencies between butterfly operations within the same stage. This parallelism is limited to $\min(n/2,1024)$-way for NTTs of size $n$, as each stage consists of $n/2$ butterflies and each thread block can support up to 1,024 threads. We refer readers to prior works~\cite{zhang2023nttongpu,zhang2023generating} on the NTTX package for further details. For BLAS operations, each CUDA thread handles the computation for one element of the input vector. Therefore, we can perfectly parallelize up to 1,024-way vector-vector BLAS operations. 
As both ZKPs and FHE require multiple NTTs and BLAS operations to run concurrently~\cite{al2022openfhe,ozcan2023homomorphic,zhang2021pipezk,kim2020accelerating}, we employ batch processing on the GPU to harness additional levels of parallelism. This allows us to comprehensively evaluate the full computational capabilities of a single GPU for executing cryptographic kernels.

\paragraph{Measuring kernel runtime.} 
We calculate the runtime for a single NTT and BLAS operation as \( t_{\text{single}} = t_{\text{all}}/{k} \), where \( k \) is the batch size. We report the steady-state runtime, defined as the minimum \( t_{\text{single}} \) achievable across all batch sizes, from one to the maximum batch size that can be compiled and run. Empirically, for NTTs, close-to-minimal steady-state runtime is achieved with a batch size greater than 8 for input bit-widths ranging from 128 to 384 bits, and greater than 64 for 768-bit inputs. For BLAS operations, close-to-minimal steady-state runtime is achieved with a batch size greater than 128.

To profile kernel runtime (i.e., $t_{\text{all}}$), we use the \verb|nsys nvprof| profiling tool provided by the NVIDIA HPC SDK from the command line. BLAS operations and smaller NTTs (less than 1,024-point) are compiled without any additional flags. For larger NTTs, we compile with \begin{footnotesize}\verb|-Xcompiler -mcmodel=medium -Xcompiler \"-Wl,--no-relax\"|\end{footnotesize} to handle very large array sizes. In line with previous work on NTT acceleration, we exclude data transfer time between the CPU and GPU from the performance measurements.

\paragraph{Hardware configuration.}
We benchmarked the performance of SPIRAL-generated cryptographic kernels across three types of GPUs from different hardware generations and price points. The NVIDIA H100 Tensor Core (H100) and NVIDIA Tesla V100 Tensor Core (V100, provided by Bridges-2 at Pittsburgh Supercomputing Center~\cite{brown2021bridges}) represent two generations of server-class GPUs, released in 2023 and 2017, respectively. The NVIDIA GeForce RTX 4090 (RTX 4090), launched in 2022, is a consumer-grade GPU primarily designed for gaming and video editing. Detailed specifications for each GPU are provided in Table~\ref{tab:gpus}.

\vspace{-2mm}
\begin{table}[!htb]
  \caption{NVIDIA GPUs used for benchmarking.}
  \vspace{-2mm}
  \centering
  \label{tab:gpus}
  \begin{tabular}{cccc}
    \toprule
    Model       & H100          & RTX 4090          & V100      \\
    \midrule
    \#Cores     & 16896         & 16384             & 5120      \\
    Max Freq.   & 1980 MHz      & 2595 MHz          & 1530 MHz  \\
    RAM Size    & 80 GB         & 24 GB             & 32 GB     \\
    Bus Type    & HBM3          & GDDR6X            & HBM2      \\
    Toolkit     & 12.2          & 12.0              & 11.7      \\                     
  \bottomrule
\end{tabular}
\vspace{-3mm}
\end{table}

\subsection{BLAS Operation Results}

We first evaluated MoMA's performance on BLAS kernels that are commonly used in encryption schemes~\cite{al2022openfhe}, namely vector multiplication, vector addition, vector subtraction, and axpy, with four different input bit-widths: 128 bits, 256 bits, 512 bits, and 1,024 bits. 
In cryptographic settings, the bit-width of the modulus is usually slightly less than a power of two or a multiple of a power of two. For example, in ZKPs, prior works~\cite{ma2023gzkp,inbasekar2024icicle} use moduli of 377 bits and 753 bits, based on the underlying elliptic curves, while for FHE, researchers~\cite{soni2023rpu} employ a 116-bit modulus. For consistency and ease of comparison, we categorize the bit-widths of prior works into the nearest (multiple of a) power-of-two category relative to their actual modulus bit-width when presenting the results for both BLAS operations and NTTs. 
In many prior approaches that use Barrett reduction for cryptographic kernels~\cite{soni2023rpu,kim2020accelerating,ozerk2022efficient,livesay2023accelerating,wang2023he}, the modulus $q$ is less than or equal to $k-4$ bits to ensure that $\mu$ remains within $k$ bits, where $k$ represents the nearest (multiple of a) power-of-two bit-width in context. Similarly, our work employs a modulus with a bit-width of $k-4$ (e.g., 124 bits in the context of 128-bit results). 
It is noteworthy that the infrastructure we developed in SPIRAL using MoMA also supports a modulus of full bit-width, employing Montgomery multiplication~\cite{montgomery1985modular}. 

We compared MoMA against the state-of-the-art integer multi-precision libraries on both CPU and GPU. The GNU Multiple Precision Arithmetic Library (GMP)~\cite{granlund1996gnu} is a widely recognized library for multi-precision arithmetic on CPUs, supporting both integer and floating-point arithmetic. GRNS~\cite{isupov2021grns}, which relies on GMP for initialization, supports basic arithmetic by using RNS to decompose very large integers into natively supported integers and employs floating point processing units on GPUs.
We benchmarked MoMA-based implementations and GRNS (version 1.1.4) on V100, and GMP (version 6.1.2) on the Intel Xeon Gold 6248 @ 2.50GHz. For each BLAS operation with each input bit-width, we executed the vector operation in batch and measured the runtime for processing \(2^{20}\) integers in total. We report the steady-state runtime per element, defined as the total runtime $t_{\text{all}}$ divided by \(2^{20}\). For the GMP-based implementation, we utilized OpenMP for parallelization on CPU, with the OpenMP directive \begin{small}\verb|#pragma omp parallel for|\end{small}.

\begin{figure*}[!htb]
\centering
    \subfloat[128-bit\label{fig:ntt_128}]{\includegraphics[width=0.88\columnwidth,trim={0 0 0 0},clip]{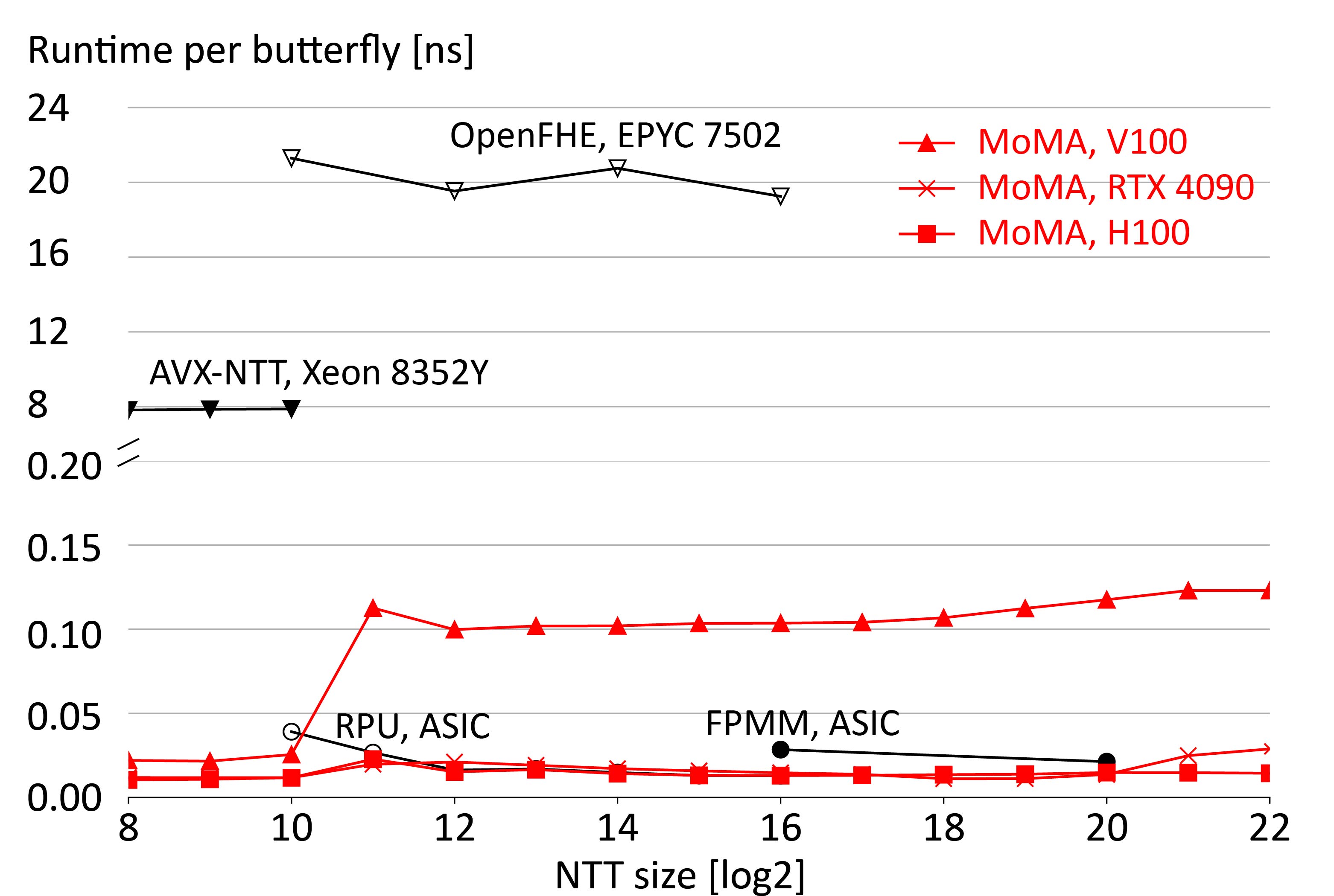}}
    \hspace{3mm}
    \subfloat[256-bit\label{fig:ntt_256}]{\includegraphics[width=0.88\columnwidth,trim={0 0 0 0},clip]{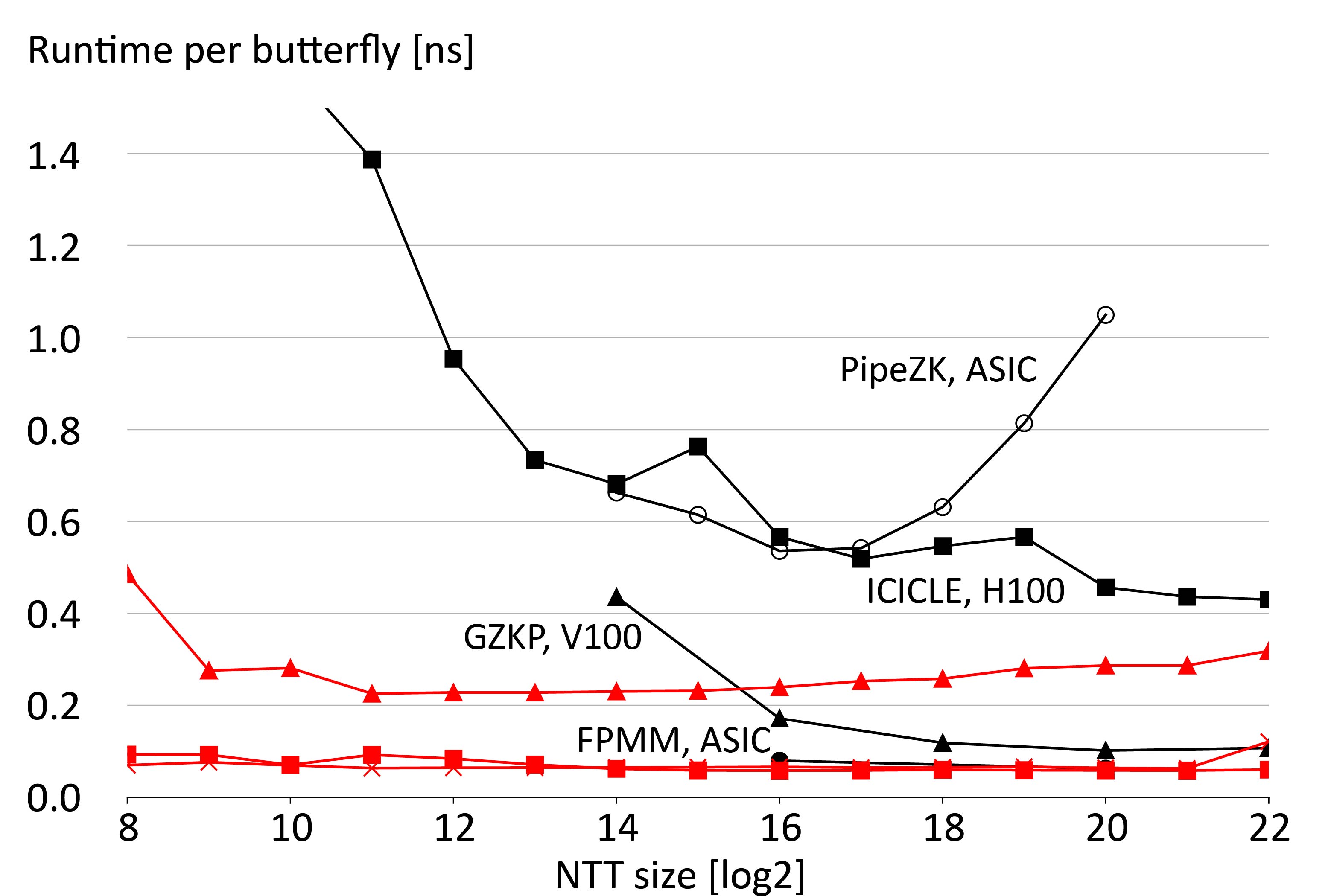}} \\ 
    \subfloat[384-bit\label{fig:ntt_384}]{\includegraphics[width=0.88\columnwidth,trim={0 0 0 0},clip]{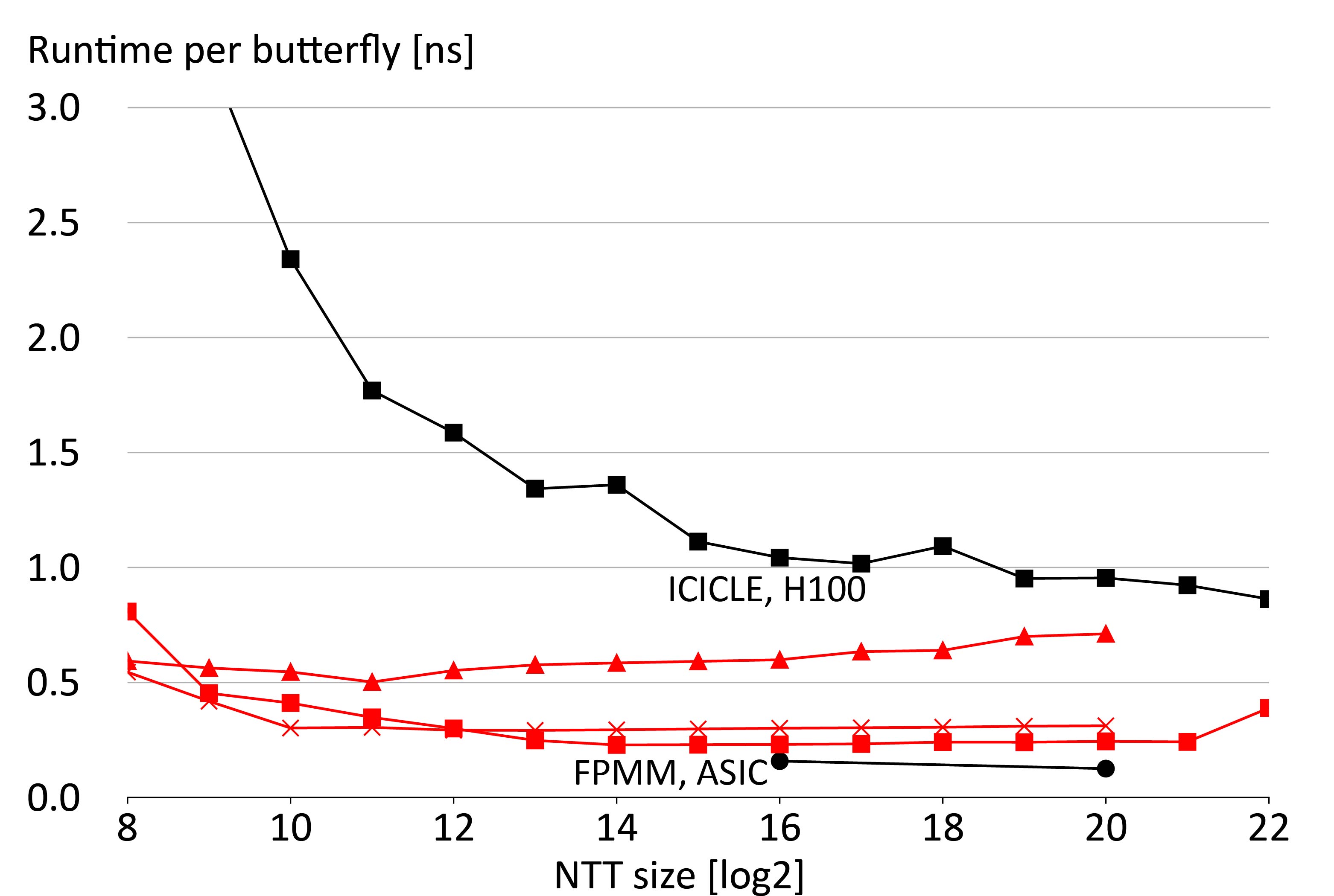}}
    \hspace{3mm}
    \subfloat[768-bit\label{fig:ntt_768}]{\includegraphics[width=0.88\columnwidth,trim={0 0 0 0},clip]{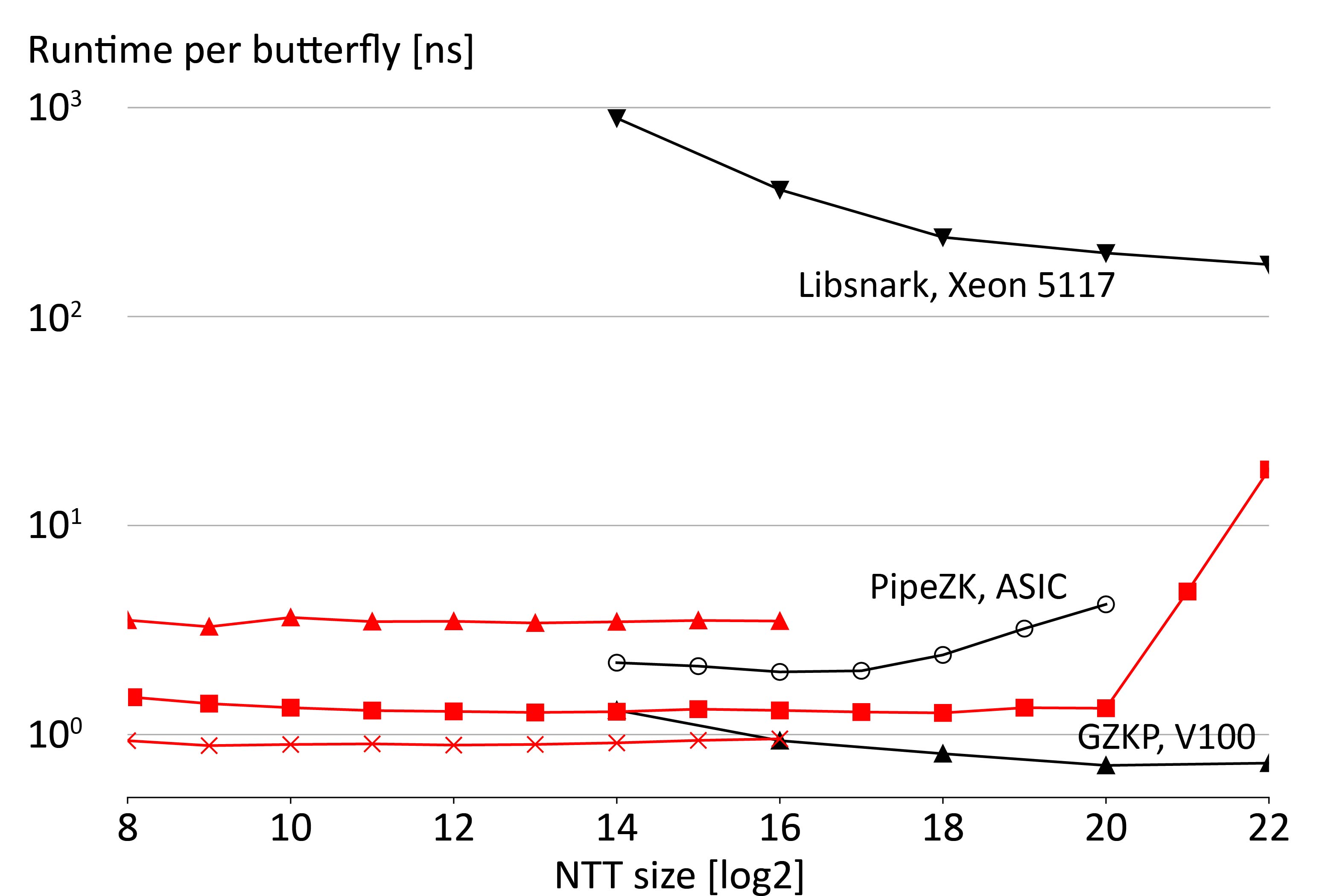}} \\ 
    \vspace{-2mm}
    \caption{Performance of NTT with various input bit-widths on CPUs, GPUs, and ASICs.}
    \label{fig:ntt_results}
    \vspace{-3mm}
\end{figure*}

The results presented in Figure~\ref{fig:blas_results} demonstrate that MoMA-based implementations outperform both GMP and GRNS across all four operations with bit-widths ranging from 128 bits to 1,024 bits, achieving speedups of at least 13 times. 
For multiplication-based kernels, namely, vector multiplication and axpy, MoMA's speedup increases relative to GRNS but diminishes compared to GMP (although still maintaining a speedup of over 10 times for 1,024-bit inputs). This behavior is expected, as GMP utilizes fast Fourier transform (FFT)-based algorithms for large bit-width multiplications, which is reflected in the fact that GMP's runtime for 512-bit and 1,024-bit inputs is lower than for 128-bit and 256-bit inputs. For addition and subtraction operations, GRNS outperforms GMP across all bit-widths, with GMP narrowing the gap as bit-width increases. MoMA achieves at least 31 times speedup over GRNS and at least 527 times speedup over GMP for addition and subtraction operations.

\subsection{NTT Results}
\label{sec:ntt_results}

\begin{figure*}[t]
\centering
    \includegraphics[width=0.9\textwidth,trim={0mm 0mm 0mm 0mm},clip]{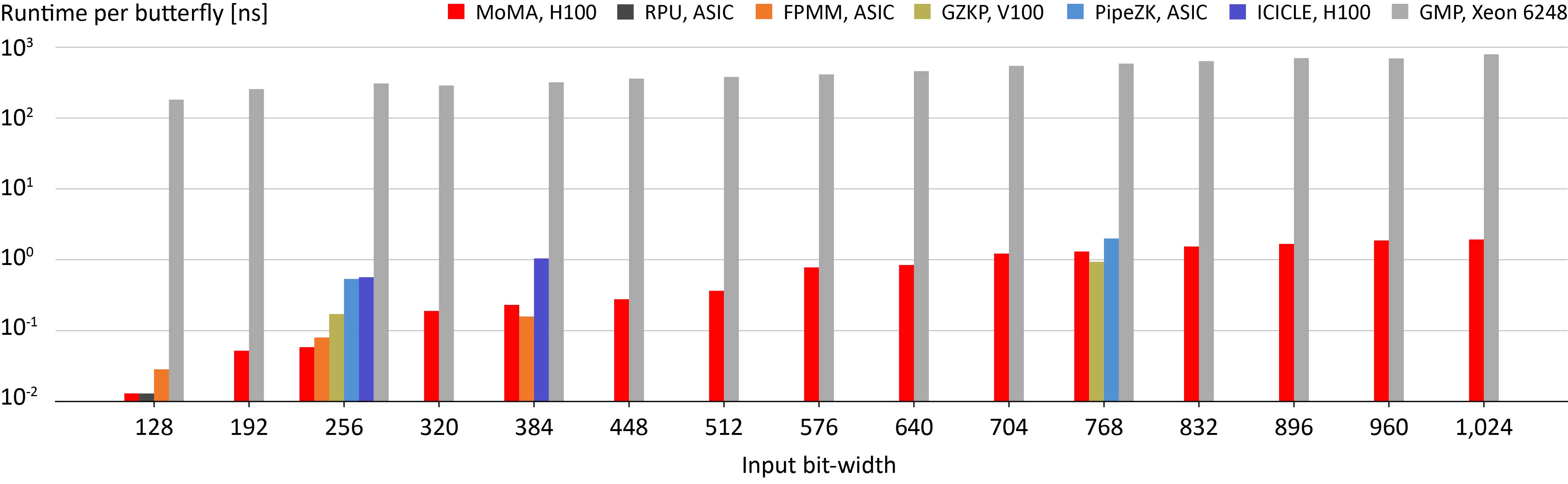}
    \vspace{-2mm}
    \caption{Performance of $2^{16}$-point NTT with input bit-widths ranging from 128 to 1,024 on CPUs, GPUs, and ASICs.}
    \label{fig:crosscut}
    \vspace{-2mm}
\end{figure*}

Following the assessment of BLAS operations, we evaluated MoMA's performance on a more complex cryptographic kernel that serves as a core component in both FHE and ZKP workloads, NTT. For NTT evaluations, we did not employ any specialized primes (e.g., Goldilocks primes~\cite{hamburg2015ed448} or Montgomery-friendly primes~\cite{bajard2021montgomery}) for performance gain, thereby ensuring the rule system's general-purpose applicability.
The code generator SPIRAL equipped with MoMA proved to be highly versatile and performant, allowing us to compare it against eight baseline implementations, each typically optimized for specific input bit-widths and NTT sizes. In terms of \textit{generalizability}, the closest comparable library is ICICLE~\cite{inbasekar2024icicle}, which supports NTTs for two input bit-widths (256 bits and 768 bits) and is applicable across all tested NTT sizes (ranging from $2^8$ to $2^{22}$). Despite this, MoMA demonstrates a significant performance advantage, achieving a 13 times speedup over ICICLE for 256-bit inputs and a 4.8 times speedup for 384-bit inputs. 
In terms of \textit{performance}, MoMA-based NTT running on RTX 4090, a \$2,000 consumer-grade GPU, outperforms NTTs on two state-of-the-art ASICs (RPU~\cite{soni2023rpu} and Zhou et al.'s work~\cite{zhou2024fully}), for 128-bit and 256-bit inputs. In the following texts, we refer to Zhou et al.'s work as FPMM and our MoMA-based NTT as MoMA. 

In the context of NTT, a butterfly operation includes one modular addition, one modular subtraction, and one modular multiplication. In Figure~\ref{fig:ntt_results}, we present the runtime per butterfly for our approach and all baselines on the y-axis. This metric is defined as $2 t_{\text{single}} / (n \log_2 n)$, where \( n \) is the NTT size, $(n \log_2 n) / 2$ is the number of butterflies in an $n$-point NTT, and $t_{\text{single}}$ is the runtime for a single $n$-point NTT. 

\paragraph{128-bit inputs.} 
In Figure~\ref{fig:ntt_128}, for NTT sizes up to $2^{10}$, the entire NTT fits within the GPU's shared memory, which accounts for the significant slowdown observed on V100 for size $2^{11}$ and larger. On H100 and RTX 4090, going out of the shared memory results in a 1.5 times slowdown. 
We compare our results against two CPU baselines, OpenFHE~\cite{al2022openfhe} (based on the benchmarking results reported by RPU~\cite{soni2023rpu}) and AVX-NTT~\cite{fu2024avxntt}, as well as two ASIC baselines, RPU and FPMM~\cite{zhou2024fully}. On H100, MoMA outperforms RPU, an accelerator specifically designed for FHE, by 1.4 times on average and FPMM by 1.8 times on average. Notably, on RTX 4090, a consumer-grade GPU, MoMA achieves an average speedup of 1.3 times over RPU and 1.7 times over FPMM. 

\paragraph{256-bit inputs.} 
Figure~\ref{fig:ntt_256} compares MoMA with two GPU-based approaches, GZKP~\cite{ma2023gzkp} and ICICLE~\cite{inbasekar2024icicle}, and two ASIC-based approaches, PipeZK~\cite{zhang2021pipezk} and FPMM. 
On H100, our approach shows a 13 times average speedup across all tested NTT sizes over ICICLE. On all three tested GPUs, MoMA outperforms PipeZK, an ASIC designed to accelerate ZKPs using a pipelined architecture. On the H100 and RTX 4090, MoMA also outperforms FPMM, an ASIC with reported results for two NTT sizes. On V100, MoMA is outperformed by GZKP for large NTT sizes due to the fact that GZKP exploits all the floating-point processing units on GPU. However, MoMA outperforms GZKP on smaller sizes, even when using only integer processing units. 

\paragraph{384-bit inputs.} 
For 384-bit inputs, as shown in Figure~\ref{fig:ntt_384}, we compare MoMA with ICICLE on H100 and FPMM, an ASIC. MoMA on H100 achieves an average speedup of 4.8 times across all tested NTT sizes against ICICLE. Moreover, on V100, MoMA also outperforms ICICLE on H100 by 3 times on average. MoMA-based NTT runs into segmentation faults at size $2^{21}$ on both RTX 4090 and V100, with an error message indicating running out of the stack space during compilation. Although MoMA outperforms FPMM for 128-bit and 256-bit inputs, FPMM achieves a 1.7 times speedup over our approach at 384-bit inputs. 

\paragraph{768-bit inputs.} 
In Figure~\ref{fig:ntt_768}, we compare MoMA with PipeZK, GZKP, and Libsnark~\cite{libsnark}. Libsnark is a CPU-based ZKP library that implements 768-bit NTTs, and we plot the benchmarking results of Libsnark as reported by GZKP. For 768-bit inputs, RTX 4090 outperforms H100 until the NTT code runs into a segmentation fault at size $2^{16}$ (which also occurs for V100). The speedup of RTX 4090 over H100 could be attributed to its higher clock speed. As the bit-width increases, each butterfly operation becomes significantly more computationally intensive. In such scenarios, a GPU with a higher clock speed, such as RTX 4090, may be more favorable. 
Both H100 and RTX 4090 outperform PipeZK, with H100 achieving a 2 times speedup over PipeZK for sizes ranging from $2^{14}$ to $2^{20}$. However, the performance of H100 degrades significantly beyond size $2^{20}$, suggesting that the hardware or compiler limits are being approached. 
For 768-bit inputs, from size \(2^{16}\) onwards, MoMA is outperformed by GZKP. This comparison highlights that utilizing all floating-point units on the GPU is a clear advantage for accelerating NTT computations, especially with large input bit-widths. 
Since MoMA rewrite rules are well-defined and composable, future efforts could leverage GPU floating-point processing units and utilize specialized primes, such as Montgomery-friendly primes for certain ZKP applications, to further improve performance.

\paragraph{Comparison among multiple input bit-widths.} 
As detailed in Section~\ref{sec:rules}, MoMA's formalism enables optimizations for inputs with non-power-of-two bit-widths. This allows MoMA to optimally support a wide range of fine-grained bit-widths, rather than resorting to zero-padding the inputs to the nearest power-of-two. Figure~\ref{fig:crosscut} illustrates this flexibility, where we compare the performance of MoMA with NTTs implemented using the state-of-the-art multi-precision library GMP, along with results from other works discussed earlier, plotted for the relevant bit-widths.
This plot can be interpreted as a cross-cut of the four subplots in Figure~\ref{fig:ntt_results}, where we use an NTT size of $2^{16}$ and assemble the subplots accordingly. We choose \(2^{16}\) as the NTT size because for this size there exists the most comparable work across multiple input bit-widths. 
The key takeaway from this plot is that MoMA demonstrates: i) significant performance improvements over libraries supporting general input bit-widths, and ii) comparable performance to approaches that utilize specialized hardware and/or are optimized for a limited set of input bit-widths. 

\begin{figure}[t]
\centering
    \subfloat[Runtime against input bit-width on H100 and RTX 4090.\label{fig:bit_width}]{\includegraphics[width=0.48\columnwidth,trim={10mm 0mm 10mm 5mm},clip]{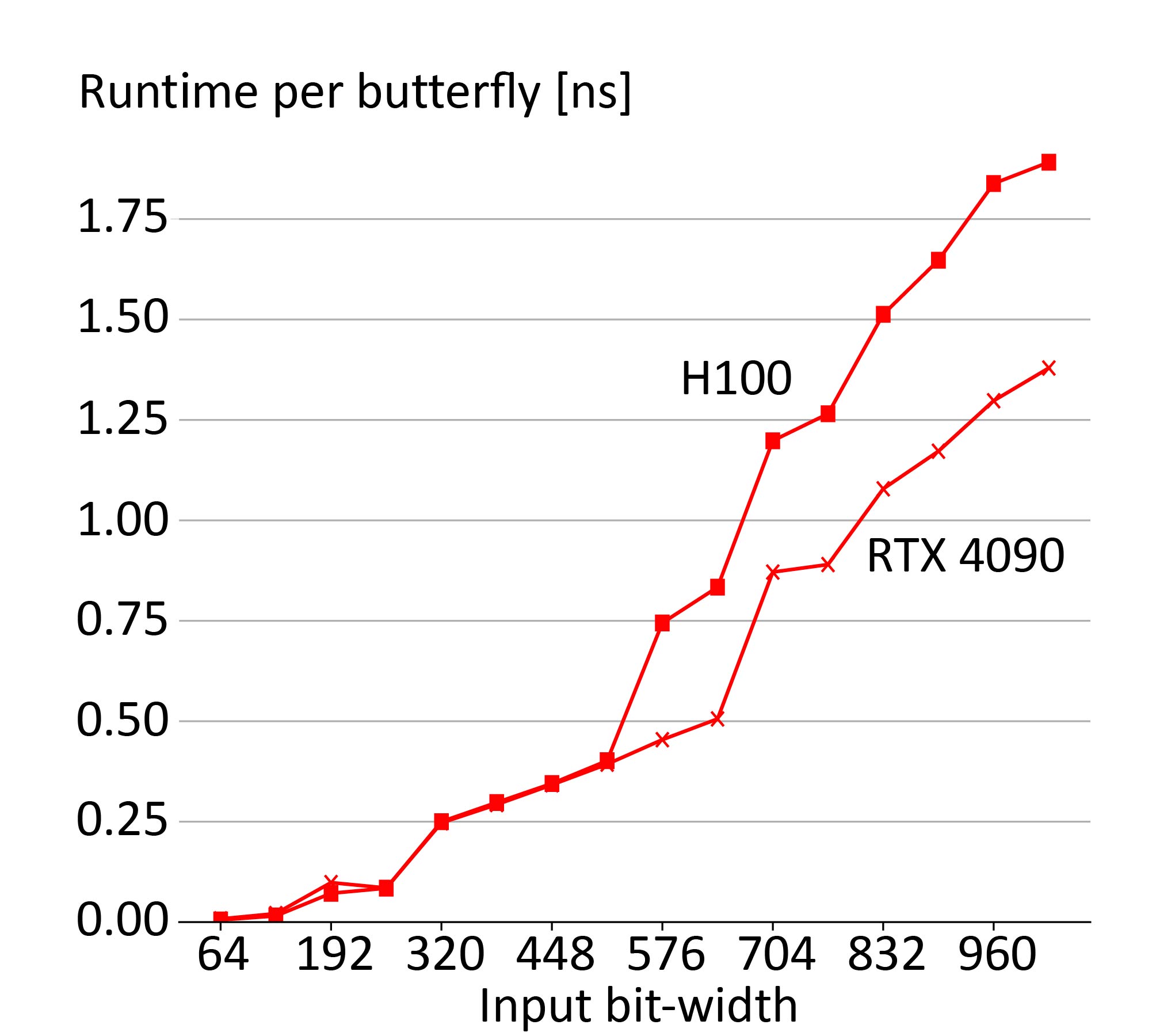}} 
    \hspace{1mm}
    \subfloat[Comparison between the Karatsuba algorithm and the Schoolbook algorithm on RTX 4090.\label{fig:mult_alg}]{\includegraphics[width=0.48\columnwidth,trim={10mm 0mm 10mm 5mm},clip]{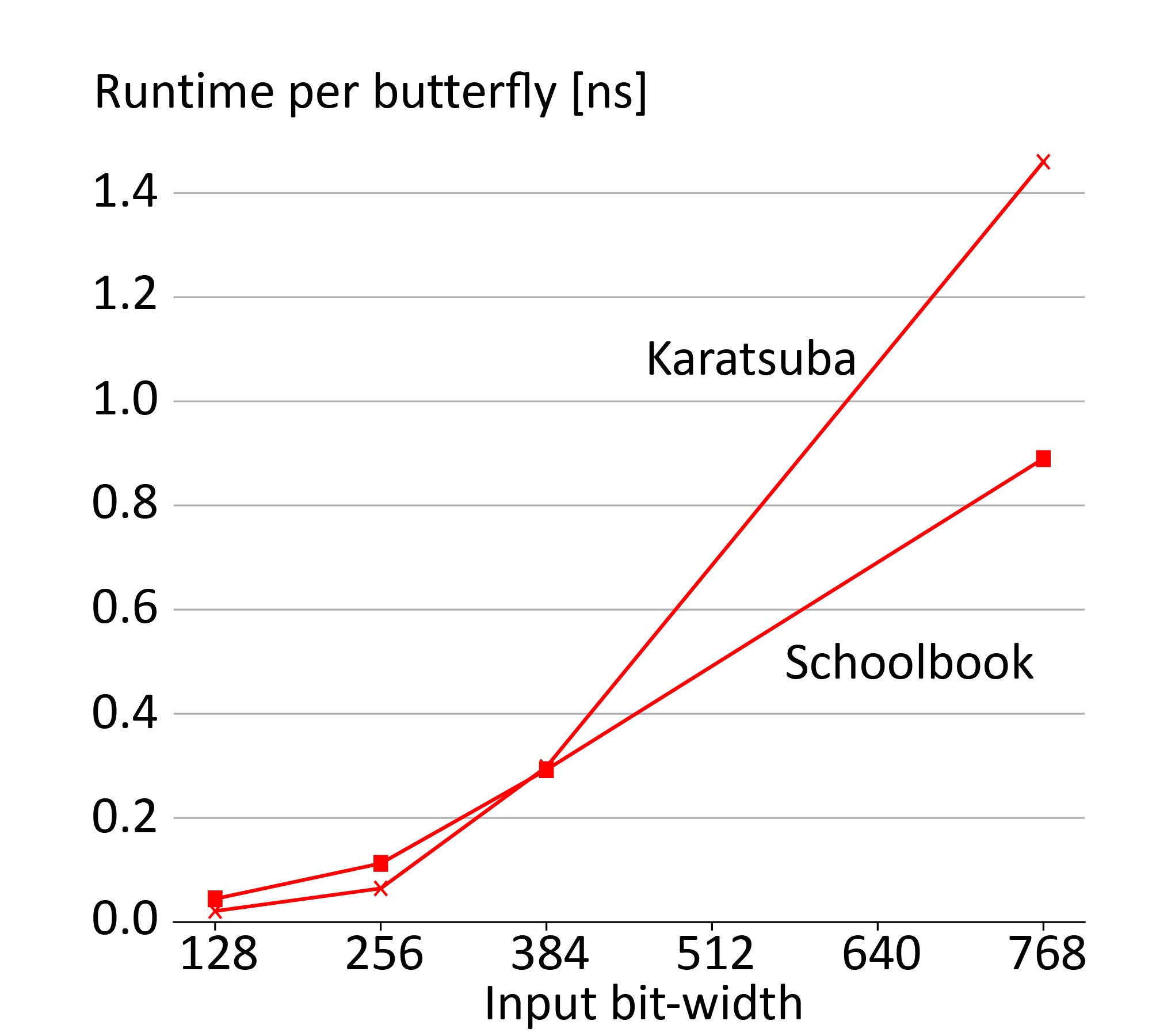}}
    \vspace{-2mm}
    \caption{Sensitivity analyses on NTT runtime.}
    \label{fig:sensitivity}
    \vspace{-1mm}
\end{figure}

\subsection{Sensitivity Analysis}

We now conduct two sensitivity analyses on MoMA's performance, examining the impact of input bit-width and the choice of multiplication algorithm on NTT runtime. Both analyses are performed with a fixed NTT size of 4,096.

\paragraph{Impact of input bit-width on runtime.}
We investigate how the performance of MoMA-based NTT scales with increasing input bit-width. In Figure~\ref{fig:bit_width}, we benchmark MoMA-based NTTs for input bit-widths ranging from 64 to 1,024 bits on both H100 and RTX 4090. We observe that both GPUs exhibit similar scaling trends with increasing bit-width. Specifically, from 320 bits to 512 bits, both RTX 4090 and H100 show a linear increase in runtime. However, at 576 bits, the runtime on H100 starts to increase non-linearly, while on RTX 4090 it maintains a linear increase up to 640 bits. Beyond 512 bits, the performance gap between RTX 4090 and H100 remains relatively constant.
On H100, we observe a 2.9 times slowdown when increasing from 64 bits to 128 bits, a 5.6 times slowdown from 128 bits to 256 bits, a 4.8 times slowdown from 256 bits to 512 bits, and a 4.7 times slowdown from 512 bits to 1,024 bits. The corresponding slowdowns on RTX 4090 are 2.7, 4, 4.6, and 3.5 times, respectively.

\paragraph{Choices of multiplication algorithm.}
As discussed in Section~\ref{sec:multi_digit}, MoMA offers two implementation choices for multiplication: i) the Schoolbook algorithm and ii) the Karatsuba algorithm. The Schoolbook algorithm for double-word multiplication requires 4 single-word multiplications and 6 single-word additions (excluding carry propagation). In contrast, the Karatsuba algorithm uses 3 single-word multiplications, 12 single-word additions/subtractions (excluding carry propagation), and several single-word comparisons. The Karatsuba algorithm reduces the number of multiplications, which are typically the most computationally intensive operations, by replacing one multiplication with several relatively cheaper additions.
In MoMA, both algorithms are included and, when implemented in the SPIRAL code generator, users can select the preferred multiplication algorithm to optimize performance on the target hardware.
Figure~\ref{fig:mult_alg} compares the performance of the Schoolbook algorithm and the Karatsuba algorithm for NTT computations on RTX 4090 across various bit-widths. For 128-bit and 256-bit inputs, the Karatsuba algorithm outperforms the Schoolbook algorithm by 2.1 times and 1.7 times, respectively. Both algorithms exhibit similar performance for 384-bit inputs, while the Schoolbook algorithm begins to outperform the Karatsuba algorithm for 768-bit inputs, with a 1.6 times speedup.

\section{Related Work}

We review the state-of-the-art approaches for multi-precision integer arithmetic and prior work on NTT acceleration.

\paragraph{Multi-precision integer arithmetic.} There are many prior work that efficiently implements multi-precision arithmetic. However, many focus on a single bit-width or a limited range of bit-widths~\cite{zhao2009implementation,zhao2010gpump,ewart2013vli,edamatsu2020accelerating,ochoa2020implementation,park2021efficient,plantard2021efficient,zhang2023nttongpu}, focus on one arithmetic operation (usually multiplication)~\cite{shieh2010word,emmart2011high,kitano2014multiple,gueron2016accelerating,dieguez2022efficient}, or is limited to a specific CPU/GPU architecture or a specific ASIC design~\cite{chang2018multiplying,soni2023rpu,rudnicki2020open,zhou2024fully}.
There are also libraries designed for generalizability, which implement arbitrary precision arithmetic. One of the most well-known and well-maintained libraries is GMP~\cite{granlund1996gnu}. Other libraries such as a library for doing number theory (NTL)~\cite{shoup2001ntl} and fast library for number theory (FLINT)~\cite{hart2013flint} also support arbitrary precision arithmetic but use GMP behind the scenes. As GMP is written in C, both GMP and libraries that depend on GMP run on CPU only. On the other hand, many programming languages, such as Python and Rust, offer native support for computations with arbitrarily large integers. However, as prior studies have shown~\cite{hughes2014mplib}, these languages are generally outperformed by GMP in terms of performance. 
GRNS~\cite{isupov2021grns} is a GPU-based arbitrary precision integer library that relies on GMP for initialization. GRNS uses RNS to break down very large integers into natively supported integers and utilizes floating-point processing units on GPU for RNS-based integer arithmetic. MoMA outperforms both GMP and GRNS on all four common input bit-width for popular encryption schemes by orders of magnitude, as shown in Figure~\ref{fig:blas_results} and~\ref{fig:crosscut}. 

\paragraph{NTT acceleration.} Numerous works have been proposed to accelerate NTT due to its significance in terms of runtime in FHE and ZKP workloads. 
Most of the prior work focuses on designing specific accelerators~\cite{samardzic2021f1,samardzic2022craterlake,soni2023rpu,wang2023sam,zhou2024fully} to address the prohibitive computational overhead introduced by FHE and ZKPs.
GPU-based approaches for accelerating NTT primarily focus on input bit-widths that is natively supported by the machine (i.e., 32 bits and 64 bits)~\cite{longa2016speeding,kim2020accelerating,ozerk2022efficient,durrani2021accelerating,shivdikar2022accelerating,wan2022novel,wang2023he,ozcan2023homomorphic,livesay2023accelerating,wang2023nttfusion}.
ICICLE~\cite{inbasekar2024icicle} is a very relevant work that offers great generalizability as a high-performance cryptographic acceleration library. As shown in Figure~\ref{fig:ntt_results}, ICICLE is the only work that can compile and run on all NTT sizes that we tested on two input bit-widths. GZKP~\cite{ma2023gzkp} is another GPU-based approach that accelerates NTT of sizes larger than machine word width. As mentioned at the end of Section~\ref{sec:ntt_results}, GZKP leverages GPU floating-point processing units, a feature that can be incorporated into MoMA given the generalizability of the rewrite rules; however, this is beyond the scope of this work.
There are many CPU-based libraries for FHE and ZKP workloads~\cite{chen2017seal,al2022openfhe,helib,chillotti2020tfhe,boemer2021intel,libsnark,bowe2020zexe}. Most of these libraries prioritize generalizability (e.g., supporting various encryption schemes) and demonstrate suboptimal performance on NTTs compared to GPU- and AISC-based approaches~\cite{soni2023rpu}. 

\section{Discussion}

As shown in Figures~\ref{fig:blas_results}, ~\ref{fig:ntt_results} and~\ref{fig:crosscut}, MoMA-based kernels demonstrate strong performance for bit-widths between 128 and 1,024 bits. However, the performance gap with the state-of-the-art multi-precision library narrows as the bit-width increases. We anticipate that, for example, GMP’s FFT-based multiplication will outperform MoMA for very large bit-widths (e.g., 8,192 bits). Nonetheless, as a rule system, MoMA can incorporate algorithms tailored to specific scenarios due to its composable and well-defined formalism. We believe that MoMA's exceptional performance on bit-widths relevant to FHE and ZKPs opens the door to exploring new encryption schemes involving large bit-widths, as the cost of exceeding machine word arithmetic has been significantly reduced.

Driven by artificial intelligence (AI) and machine learning (ML) applications, customized hardware \sloppy{continues trending toward smaller machine word widths} (e.g., 16-bit unsigned integer on the Cerebras Wafer Scale Engine~\cite{selig2022cerebras}). However, FHE, an enabling encryption scheme for privacy-preserving AI/ML, requires support for large integer arithmetic. FHE-specific hardware is therefore designed to natively support large bit-width arithmetic operations (e.g., 128-bit modular arithmetic supported by RPU~\cite{soni2023rpu}). 
As future work, we plan to investigate whether MoMA can work effectively on AI/ML-specialized hardware to keep the runtime of FHE-based workloads manageable with native small integer arithmetic. In other words, we aim to explore whether MoMA can bridge the gap between these two conflicting hardware trends at the software level. 
Moreover, we plan to integrate MoMA into various compilers, ranging from generic frameworks like LLVM to specialized compilers such as the CSL compiler for Cerebras accelerators, further expanding the impact of our approach.

\section{Conclusion}

To address the critical need for efficient large integer arithmetic in cryptographic applications, our work formally defines MoMA, which decomposes large bit-width integer arithmetic into operations based on machine words. We developed a mathematically formal rewrite system that implements MoMA and can be seamlessly integrated into compilers and code generators. 
For evaluation, we implemented the MoMA rule system in SPIRAL and generated cryptographic kernels, including BLAS operations and NTTs, across three types of GPUs. Our results show that the generated BLAS operations outperform state-of-the-art multi-precision libraries by several orders of magnitude, while MoMA-based NTT achieves near-ASIC performance on commodity GPUs.


\begin{acks}
This material is based upon work supported by the National Science Foundation under Grant No. 1127353, the U.S. Department of Energy, Office of Science, Office of Advanced Scientific Computing Research under Award Number DE-FOA-0002460, and PRISM, a center in JUMP 2.0, a Semiconductor Research Corporation (SRC) program sponsored by the Defense Advanced Research Projects Agency (DARPA). 
This work used Bridges-2 at Pittsburgh Supercomputing Center through allocation CIE160039 from the Advanced Cyberinfrastructure Coordination Ecosystem: Services \& Support (ACCESS) program, which is supported by National Science Foundation grants \#2138259, \#2138286, \#2138307, \#2137603, and \#2138296.
Any opinions, findings, and conclusions or recommendations expressed in this material are those of the authors and do not necessarily reflect the views of the National Science Foundation, the U.S. Department of Energy, and DARPA.
Franz Franchetti was partially supported as the Kavčić-Moura Professor of Electrical and Computer Engineering. 
\end{acks}


\appendix
\section{Artifact Appendix}

\subsection{Abstract}


Our artifact~\cite{zhang2024artifact} includes the source code for MoMA, requiring NVIDIA GPUs along with \verb|nvcc| and \verb|nsys nvprof| for compilation and performance profiling. While installing the SPIRAL code generation system is highly recommended for full reproducibility and customization, it is not required, as we provide pre-generated NTT and BLAS code from SPIRAL to reproduce key results.

\subsection{Artifact Checklist (Meta-information)}


{\small
\begin{itemize}
  \item {\bf Program: } MoMA as part of the SPIRAL NTTX package. The artifact includes benchmarking data for NTT and BLAS operations.
  \item {\bf Compilation: } The artifact requires \verb|nvcc| >= 11.7 to compile the generated CUDA code and \verb|nsys| >= 2022.4.2 for performance measurement.
  \item {\bf Transformations: } SPIRAL >= 8.5.0 is required for code generation; however, pre-generated code is provided for users who prefer not to install SPIRAL or spend time on code generation.
  \item {\bf Hardware: } NVIDIA GPUs as detailed in Table~\ref{tab:gpus}.
  \item {\bf Execution: } We provide a Bash script to benchmark the pre-generated code directly or to invoke SPIRAL for code generation followed by benchmarking.
  \item {\bf Metrics: } Execution time.
  \item {\bf Output: } Performance measurements will be displayed in the terminal window. 
  \item {\bf Experiments: } A detailed \verb|README| file in the \verb|nttx| directory explains how to reproduce key results, with or without SPIRAL installation. Expected outputs for specific cases are also provided in the \verb|README| file.
  \item {\bf How much disk space required (approximately)?: } Around 50 MB.
  \item {\bf How much time is needed to prepare workflow (approximately)?: } Using pre-generated code, the user can immediately execute the workflow via an automated benchmarking script. Installing SPIRAL typically takes less than an hour.
  \item {\bf How much time is needed to complete experiments (approximately)?: } Using pre-generated code, benchmarking 1,024-point NTT with 256-bit inputs takes approximately five minutes. Starting from code generation, this process takes approximately six minutes. Completing all experiments on all three GPUs from end to end may take days, as the code generation time increases exponentially with the input bit-width. 
  \item {\bf Publicly available?: } Yes. 
  \item {\bf Archived (provide DOI)?: } \href{https://doi.org/10.5281/zenodo.14564393}{10.5281/zenodo.14564393}.
\end{itemize}

\subsection{Description}

Here we provide a short description of how the artifact is delivered and its dependencies. 

\paragraph{How delivered.}
%
The artifact is provided as a zip file available on Zenodo\footnote{\url{https://doi.org/10.5281/zenodo.14564393}} and as a repository on GitHub\footnote{\url{https://github.com/naifeng/moma}}. The artifact requires approximately 50 MB of disk space.

\paragraph{Hardware dependencies.} NVIDIA GPUs. The key results in this paper were obtained using NVIDIA H100, V100, and RTX 4090 GPUs, with details provided in Table~\ref{tab:gpus}. 

\paragraph{Software dependencies.} \verb|nvcc| >= 11.7 and \verb|nsys| >= 2022.4.2. SPIRAL >= 8.5.0 is highly recommended but not required.


\subsection{Installation}

SPIRAL can be installed following its installation guide\footnote{\url{https://github.com/spiral-software/spiral-software}}. To link the artifact with SPIRAL, place the artifact into the \verb|spiral-software/namespaces/packages/| subdirectory of the SPIRAL installation tree.

\subsection{Experiment Workflow}

A detailed \verb|README| file is provided at \verb|nttx/README.md|. For example, with SPIRAL installed, users can reproduce our 128-bit NTT results on H100 by running the following commands:
\begin{minted}[frame=none,obeytabs=true,tabsize=4,fontsize=\footnotesize,bgcolor=bg]{text} 
$ cd nttx/cuda/cuda-test
$ bash ./benchmark.sh -d 128 -p h100
\end{minted}
The option \verb|-d| specifies the input bit-width (128, 256, 384, or 768), while \verb|-p| enables performance tuning for the target platform (currently supported platforms are H100, V100, and RTX 4090). For other platforms, the \verb|-p| option can be omitted, which defaults to \verb|-p general|.

\subsection{Evaluation and Expected Result}

After completing the experiments, the terminal will display the results as follows: for NTT, the runtime per butterfly and the runtime per NTT for each NTT size; for BLAS operation, the runtime per element and the runtime per vector operation. All runtimes are reported in nanoseconds.






\bibliographystyle{ACM-Reference-Format}
\bibliography{ref}

\end{document}